\documentclass[11pt]{article}
\pdfoutput=1
\usepackage{jcappub,natbib,float,caption,comment,bm,epsfig}
\usepackage{graphicx}
\usepackage[dvipsnames]{xcolor}
\usepackage[utf8]{inputenc}
\usepackage{color, colortbl}
\definecolor{Gray}{gray}{0.9}
\usepackage[first=0,last=9]{lcg}

\usepackage{float}

\usepackage[normalem]{ulem}
\usepackage{cancel}

\usepackage{slashed}
\usepackage{graphicx}
\usepackage{subfig}

\newcommand{\beq}{\begin{equation}}
\newcommand{\eeq}{\end{equation}}
\newcommand{\bea}{\begin{eqnarray}}
\newcommand{\eea}{\end{eqnarray}}

\newcommand\lsim{\mathrel{\rlap{\lower4pt\hbox{\hskip1pt$\sim$}}
        \raise1pt\hbox{$<$}}}
\newcommand\gsim{\mathrel{\rlap{\lower4pt\hbox{\hskip1pt$\sim$}}
        \raise1pt\hbox{$>$}}}

\title{\center{Radio signatures from encounters between Neutron Stars and QCD-Axion Minihalos around Primordial Black Holes}}
\author[a,b]{Sami Nurmi,}
\author[a,b]{Enrico D. Schiappacasse,}
\author[c]{and Tsutomu T. Yanagida}
\affiliation[a\,]{Department of Physics, P.O. Box 35 (YFL),
FI-40014 University of Jyv$\ddot{\text{a}}$skyla, Finland}
\affiliation[b\,]{Helsinki Institute of Physics, P.O. Box 64, FIN-00014 University of Helsinki,  Finland}
\affiliation[c\,]{Tsung-Dao Lee Institute $\&$ School of Physics and Astronomy, Shanghai Jiao Tong University, 
 200240 Shanghai, China}

\emailAdd{sami.t.nurmi@jyu.fi}
\emailAdd{edschiap@uc.cl}
\emailAdd{tsutomu.tyanagida@ipmu.jp}

\abstract{
Probing the QCD axion dark matter (DM) hypothesis is extremely 
challenging as the axion interacts very weakly with Standard Model particles.
We propose a new avenue to test the QCD axion DM via transient radio signatures coming from encounters between neutron stars (NSs) and axion
minihalos around primordial black holes (PBHs). We consider a general QCD axion scenario in which the PQ symmetry breaking occurs 
before (or during) inflation coexisting with a small fraction of DM in the form of PBHs. The PBHs will unavoidably acquire around them axion minihalos with the typical length scale of parsecs. The axion density in the minihalos may be much higher than the local DM density, and the presence of these compact objects in the Milky Way today provides a novel chance for testing the axion DM hypothesis. We study the evolution of the minihalo mass distribution in the Galaxy accounting for tidal forces and estimate the encounter rate between NSs and the dressed PBHs.
 We find that the encounters give rise to transient line-like emission of radio frequency photons produced by the resonant axion-photon conversion in the NS magnetosphere and the characteristic  signal could be detectable with the sensitivity of current and prospective radio telescopes. \textcolor{black}{ It would be important to investigate in detail search strategies for such signals which would provide a novel pathway for QCD axion detection.} 
 }

\begin{document}

\maketitle
\flushbottom
%\tableofcontents

%\newpage

\section{Introduction}
\label{sec:section1}
In this paper, we propose for the first time a new avenue for 
detecting the QCD axion via transient 
radio signatures coming from the encounters between neutron stars (NSs) 
and axion dark halos around primordial black holes (PBHs). 
 Crossing of the NS magnetosphere through the axion minihalo 
leads to monochromatic radio signals
due to axion-photon resonant conversion~\cite{Pshirkov:2007st}. 
We show that the signal should
be detectable on the Earth under suitable conditions.

Even though a wide range
of observations coming from, for example, gravitational lensing~\cite{1987ARA&A..25..425T}, hot gas in clusters, the cosmic microwave background radiation~\cite{Ade:2015xua, Aghanim:2018eyx} and the hierarchical structure formation of the Universe~\cite{Springel:2005nw}, are 
well explained by the inclusion of cold dark matter, its
origin in the particle physics context 
remains unknown~\cite{Bertone:2016nfn,Freese:2017idy}. The strong CP problem~\cite{PhysRevLett.38.1440, PhysRevLett.40.223, PhysRevLett.40.279} and unification with gravity in the context of string theory make the QCD axion~\cite{DiLuzio:2020wdo} one of the strongest DM candidates. In last years, the quest for the axion dark matter has strongly attracted the attention of the physics community leading to a diverse
search program including, for example, haloscopes~\cite{PhysRevD.42.1297, PhysRevD.64.092003, PhysRevLett.120.151301, Zhong:2018rsr}, helioscopes~\cite{Anastassopoulos:2017ftl} and indirect axion searches~\cite{Iwazaki:2014wka, Iwazaki:2017rtb, Hertzberg:2018zte, Huang:2018lxq, Hook:2018iia, Safdi:2018oeu, Hertzberg:2020dbk, Edwards:2020afl}.  Interesting enough, most of these searches are based on the axion-photon coupling through the interacting Lagrangian $\Delta \mathcal{L} \sim g_{a\gamma\gamma} \phi\, \bf{E}\cdot \bf{B}$, where $\phi$ is the axion field, $g_{a\gamma\gamma}$ is the axion-photon coupling constant, and $\bf{E}$ and $\bf{B}$ are the electromagnetic field components, respectively. 

The axion searches currently place very little constraints on the QCD axion DM and it is therefore extremely important to study other indirect ways to test this theoretically well motivated setup. 
In this paper, we show that the presence of 
a small abundance of primordial black holes, which form a subdominant co-dark matter component,
enhances the probability for detecting the QCD axion via indirect searches. 
It should be noted that
the LIGO-Virgo gravitational wave detection~\cite{Abbott:2016blz, Sasaki:2016jop} and recent NANOGrav results~\cite{Arzoumanian:2020vkk, Kohri:2020qqd, Inomata:2020xad} 
 may indeed suggest the existence of a small fraction of DM in PBHs in the $\mathcal{O}(1-10)M_{\odot}$ mass range.
PBHs are one of the oldest DM candidate~\cite{1971MNRAS.152...75H, 1974MNRAS.168..399C, 1975Natur.253..251C, Khlopov:2008qy} and they have been studied in several setups with other DM candidates such as the axion~\cite{PhysRevLett.122.101301, Hertzberg:2020hsz} and  WIMPs~\cite{Scott:2009tu, Adamek:2019gns, Hertzberg:2020kpm}.

We consider a general QCD axion model in which the PQ symmetry is broken 
before or during inflation. Our results may apply also in the case where the PQ symmetry is broken after inflation, provided that the domain wall number $N_{\text{DW}} =1$
so that the well known domain wall problem is avoided.  When the Universe temperature drops to the QCD scale, the QCD instanton effect lifts the axion potential so that the axion acquires mass. Then, the axion field starts to roll down to one of the $N_{\text{DW}}$ degenerate minima so that domain walls are formed keeping away the
different vacua. If $N_{\text{DW}}$ = 1, axionic strings formed during the PQ symmetry breaking  attach to each domain wall forming  disk-like objects which quickly collapse under their surface tension~\cite{PhysRevLett.48.1867}.  This phenomenon leads to an upper bound for the axion decay constant to avoid an overabundance of DM such that 
$F_a \lesssim (4.6$–$7.2)\times 10^{10}\,\text{GeV}$~\cite{Kawasaki:2014sqa} . For $N_{\text{DW}} > 1$, the string-wall network is long-lived and the QCD axion is excluded in the standard scenario~\footnote{Although it may be rescued by including a bias term in the potential of the PQ field as shown in Eq.(2.40) of Ref.~\cite{Kawasaki:2014sqa}.}. 
If the PQ symmetry is broken before or during inflation, the axion field acquires quantum fluctuations proportional to the Hubble parameter during inflation leading to the so-called isocurvature perturbation problem~\cite{AXENIDES1983178, PhysRevD.32.3178}. However, several possible solutions to this issue have been already presented in the literature (e.g., see Ref.~\cite{Kawasaki:2013iha}). 
Since inflation expands a tiny region, where the axion field is homogeneous, to a region larger than the current Hubble horizon, the entire observable Universe holds the same value for the axion field. Thus, there are no production of domain walls as before and the axion abundance is given by the initial misalignment angle $\theta_i$. For $\theta_i \sim 1$,
the upper bound over the axion decay constant to avoid the overclosure of the Universe read as $F_a \lesssim 10^{12}\, \text{GeV}$ ~\cite{PRESKILL1983127, ABBOTT1983133, DINE1983137}~\footnote{
However, if we require a small $\theta_i$ in the context of the axion anthropic window~\cite{PhysRevLett.52.1725, LINDE199138, Wilczek:2004cr, PhysRevD.73.023505}, there is no upper bound on the axion decay constant from axion overabundance.}.

For typical QCD axion models, we have the Kim-Shifman-Vainshtein-Zakharov (KSVZ)~\cite{PhysRevLett.43.103} and the Dine-Fischler-Srednicki-Zhitnitsky (DFSZ)~\cite{DINE1981199, Zhitnitsky:1980tq} models. In the KSVZ model the number of degenarate minima in the QCD potential is the number of heavy quarks which carry U(1) PQ charge, so that $N_{\rm DW} = 1$ may be realized. In the DFSZ model, the domain wall number 
equals or is twice the number of flavours
of quarks carrying U(1) PQ charge, namely $N_{\text{DW}} = 3\,\, \text{or}\,\, 6$.
Apart from the phenomenological point of view, there is no general theoretical reason to choose one model over the other.

The QCD axion mass is related to the axion decay constant as~\cite{PhysRevLett.40.223}
\begin{equation}
 m_a = \frac{f_{\pi}m_{\pi}\sqrt{m_u m_d}}{F_a\left( m_u+m_d \right)} \approx 10^{-5}\,\text{eV} \left( \frac{6\times10^{11}\,\text{GeV}}{F_a} \right)\,,    
\end{equation}
where $f_{\pi}$ and $m_{\pi}$ are the pion decay constant and mass, respectively, and $m_{u}$ and $m_d$ are the up and down quark masses, respectively. Even though the axion interactions with Standard Model particles scale as $\sim f_{\pi}/F_a$, there are several non-negligible model-dependent factors. For the case of our interest, the axion-photon coupling constant is given in terms of the electromagnetic (E) and color (N) anomalies of the axial current associated with the axion as follows~\cite{diCortona:2015ldu}
\begin{equation}
g_{a\gamma\gamma} = \frac{\alpha_{\text{em}}}{2\pi F_a} \left(  \frac{E}{N}-1.92\right) = \left( 0.203 \frac{E}{N} - 0.39 \right) \left(\frac{m_a}{\text{GeV}}\right) \text{GeV}^{-1}\,,    
\end{equation}
where $\alpha_{\text{em}}$ is the fine structure constant. The dependence of the axion-photon coupling on the axion decay constant and the ratio $E/N$ makes $g_{a\gamma\gamma}$ varies several orders of magnitude. In the present work, we will consider as benchmark models the DFSZ and KSVZ models in which $E/N$ is equal to $8/3$ and $0$, respectively, so that
\begin{align}
g^{\text{DFSZ}}_{a\gamma\gamma} &\simeq  9\times 10^{-14}\, \text{GeV}^{-1} \left(\frac{10^{10}\,\text{GeV}}{F_a} \right)\,,\\
|g^{\text{KSVZ}}_{a\gamma\gamma}| &\simeq  2\times 10^{-13}\, \text{GeV}^{-1} \left(\frac{10^{10}\,\text{GeV}}{F_a} \right)\,.
\end{align}
We will take the axion mass to be a free parameter but within the range defined by the upper bound over the axion decay constant of the classical QCD axion window and the largest axion mass at which the resonant axion-photon conversion takes place in the NS magnetosphere. In the parameter space of our interest, we will see this mass range is approximately equivalent to $10^{10} \,\text{GeV} \lesssim F_a \lesssim 10^{12}\, \text{GeV}$ in terms of the axion decay constant. 

The setup we are studying, in which the axion is the dominant DM component and PBHs constitute a subdominant DM component, leads to the presence of PBHs surrounded by axion minihalos in the Milky Way today. These astrophysical objects are sometimes called dressed PBHs in the literature and we will also use this terminology here.
According to the theory of spherical gravitational collapse~\cite{1985ApJS...58...39B}, any overdensity in the DM distribution will seed the growth of a minihalo. Thus, the smooth axion DM background will be accumulated around PBHs mainly during the matter-dominated era, so that dressed PBHs will 
end up
in galactic halos. Generally speaking, probing the axion DM hypothesis is notably hard since the QCD axion is predicted to interact very weakly with Standard Model particles. That is why the presence of  
dressed PBHs, where the axion density may reach several orders of magnitude larger than its local DM value, offers an unique opportunity for axion DM detection. The resonant axion-photon conversion in a cosmological setup was  discussed for the first time by one of us in Ref.~\cite{Yanagida:1987nf}. 
Two decades later, the same resonant phenomenon was studied in  Ref.~\cite{Pshirkov:2007st} in the neutron star environment as a novel way of testing the axion hypothesis. This work was extended later in Refs.~\cite{Hook:2018iia, Huang:2018lxq}, where the sensitivity of current and prospective telescopes with respect to the associated radio signals was analyzed in detail. A three-dimensional computation of the photon flux, considering the axion phase-space distribution and the particular configuration of the NS magnetosphere, was performed in Ref.~\cite{Leroy:2019ghm}. In addition, the axion-photon conversion has been also studied beyond the DM background scenario by considering axion DM substructures such as axion stars~\footnote{An axion star or axion DM clump is a particular type of boson star (for a review about
boson stars see Ref.~\cite{Liebling2012} and for novel extensions see Refs.~\cite{Choi:2019mva, Horvat:2012aq}), which is a
self-gravitating bound state of a Bose-Einstein condensate of axions~\cite{Guth:2014hsa, Schiappacasse:2017ham, Hertzberg:2018lmt, Hertzberg:2018zte, Hertzberg:2020dbk, Visinelli:2017ooc}.}~\cite{Iwazaki:2014wta, Iwazaki:2017rtb, Buckley:2020fmh}, axion spikes around intermediate mass black holes~\cite{Edwards:2019tzf},  and axion miniclusters~\cite{Edwards:2020afl}. %Since the typical length and mass scales of QCD axion stars are the order of $10^2\,\text{km}$ and $10^{-11} M_{\odot}$~\cite{Schiappacasse:2017ham}, respectively, the encounter rate between them and NSs is highly suppressed. 
Note that the axion miniclusters ~\cite{Hogan:1988mp, Kolb:1993zz, Kolb:1993hw, Kolb:1994fi} may be formed only if 
the PQ symmetry is broken after inflation and if the QCD axion potential has just one minimum, according to numerical simulations performed in Ref.~\cite{Kawasaki:2014sqa} as explained before. Our interest is to analyze the photon emission from axion resonant conversion by encounters between NSs and dressed PBHs in the Galactic halo. Considering a general QCD scenario in which the PQ symmetry is broken before (or even during) inflation, if a small fraction of DM in PBHs exists, as suggested by LIGO-Virgo observations~\cite{Abbott:2016blz, Sasaki:2016jop} and NANOGrav results~\cite{Arzoumanian:2020vkk, Kohri:2020qqd, Inomata:2020xad}, then the formation of dressed PBHs will take place. Since their typical length scale is the order of parsecs and the axion density within their minihalos reaches values much higher than the local DM density, these astrophysical compact objects provide a unique chance for testing once and for all the axion DM hypothesis.

The present paper is organized as follows. In Sec. 2, we briefly review the DM minihalo formation around PBHs when they constitute a small initial fraction of the total DM content. We make special emphasis on the DM radial profile of these minihalos and propose a reasonable initial mass distribution for dressed PBHs immediately after being incorporated to galactic halos. In Sec. 3, we estimate the effects of tidal forces existing in the Milky Way Galaxy acting on dressed PBHs. In Sec. 4, we estimate the encounter rate between NSs and dressed PBHs in terms of their location in the galaxy. In Sec. 5, we estimate the transient radio signal during a NS-dressed PBH encounter due to the axion-photon resonant conversion in the NS magnetosphere. We also estimate the sensitivity of current and prospective radio telescopes for the QCD axion detection. Finally, in Sec. 6, we discuss and summarize the main ideas of the paper.     
\section{Axion Dark Minihalos around PBHs}
\label{dminihalo}

Primordial black holes which are formed with masses larger than 
$\sim10^{15}$ gram do not evaporate but begin to acquire a dark matter halo by gravitationally capturing the surrounding
axion DM from the smooth background.  

Since PBHs are local overdensities in the
axion DM distribution, they inevitable seed the growth of spherically
symmetric minihalos. The spherical gravitational collapse (or secondary infall) theory
tells us that at about the time of radiation-matter equality, $t\sim t_{\text{eq}}$, self-similar dark halos begins to grow around PBHs, during the matter dominated era proportional to the cosmological time as $t^{2/3}$~\cite{1985ApJS...58...39B}. Assuming the absence of tidal forces over PBHs, negligible peculiar velocities and an initial dark matter background in the Hubble flow,
the virialized 
minihalo mass and radius scale
with the redshift as~\cite{Mack:2006gz, Ricotti:2007au}
\begin{align}
M_{\textrm{halo}} (z) &= 3\left(\frac{1000}{1+z}\right) M_{\textrm{PBH}}\,,\label{mh}\\
R_{\textrm{halo}} (z) = 0.019\, &\textrm{pc} \left( \frac{M_{\textrm{halo}}}{M_{\odot}} \right)^{1/3} \left( \frac{1000}{1+z} \right)\,,\label{rh}
\end{align}
where the minihalo radius is about one third of the turnaround radius or
radial distance at which the DM shell was able to break free from the Hubble flow. Here we need to mention that Eq.~(\ref{mh}) should be
considered as an optimistic upper bound to the mass growth of minihalos.
Indeed, 
the assumption of isolated and stationary PBHs made here 
begins to break down around $z\sim (30-10)$, e.g. at the time of first galaxies formation.

The density profile of the internal structure of these minihalos can be readily derived from Eqs.~(\ref{mh}) and~(\ref{rh}) as
%\end{linenumbers}
\begin{equation}
\rho_{\textrm{halo}}(r) = \frac{1}{4\pi r^2} \frac{dM_{\textrm{halo}}(r)}{dr}
%\simeq 2.7\,M_{\odot}{\textrm{pc}}^{-3}\left( \frac{\textrm{pc}}{r} \right)^{9/4} \left( \frac{M_{\textrm{PBH}}}{\,M_{\odot}}  \right)^{3/4}\,,\label{rho}
\simeq 0.23\,M_{\odot}{\textrm{pc}}^{-3}\left( \frac{R_{\text{halo}}}{r} \right)^{9/4} \left( \frac{10^2 M_{\textrm{PBH}}}{\,M_{\text{halo}}}  \right)^{3}\,,\label{rho}
\end{equation}
%\begin{linenumbers}
which is valid for distances at which $M_{\text{halo}}(r) > M_{\text{PBH}}$. This profile agrees up to a numerical factor of two with the more recent work performed in Ref.~\cite{Berezinsky:2013fxa}.  The $\rho_{\text{halo}}\sim r^{-9/4}$ density profile was confirmed by the first N-body simulations of a universe containing a subdominant fraction of PBHs in a background of DM particles performed in Ref.~\cite{Adamek:2019gns}. 
One crucial question to be answered is the total final mass of the induced minihalo around the central PBH.  Taking into account the presence of the cosmological constant, there is a \textit{last bound shell}, namely, a last shell of DM which may be accreted by the black hole. All shells with a radius larger than that particular shell will unavoidably continue expanding in the Hubble flow. Based on that, the maximum minihalo mass is estimated to be $M_{\text{halo}}\approx 1500\, M_{\text{PBH}}$ in Ref.~\cite{Mack:2006gz} by calculating the radial infall of a DM shell under the effect of the cosmological constant. However, the minihalo mass probably will not reach this final mass since dressed PBHs begin to interact among them and/or run out of matter to accrete as minihalos grow. This scenario breaks the assumptions of isolated and stationary PBHs embedded in a  homogeneous background
made in the secondary infall theory. The final induced minihalo mass is
calculated in Ref.~\cite{Berezinsky:2013fxa} by assuming that the induced minihalo stops the accretion process in the nonlinear regime when density perturbations around dressed PBHs are the order of the growing minihalo mass. Under this consideration, the final average mass for minihalos is reported to be about $M_{\text{halo}} \sim (10^{1.5}-10^{2.5})M_{\text{PBH}}$ for the range $M_{\text{PBH}} \sim (10^{-8}-10^2)\,M_{\odot}$ (see Eq. (35) in Ref.~\cite{Berezinsky:2013fxa}).

At around $z \sim 6$, dressed PBHs will be incorporated to galactic
halos 
and the mass-radius relation of the minihalos
is readily derived from Eqs.~(\ref{mh}) and ~(\ref{rh}) as
\begin{equation}
M_{\text{halo}}(R_{\text{halo}}) = 3^{3/4} M_{\text{PBH}}\left( \frac{R_{\text{halo}}}{0.019\, \text{pc}} \right)^{3/4} \left( \frac{M_{\odot}}{M_{\text{PBH}}} \right)^{1/4}\,. \label{mr}   
\end{equation}
We do not expect a monochromatic mass function for the minihalo mass in terms of the central PBH mass but certain spread of masses around an average value. We model the initial mass function (IMF) before any disruption effects by a Gaussian distribution with an average minihalo mass $\overline{M}_{\text{halo,0}}$ and variance  $\sigma_{\text{halo,0}}$. We assume the distribution is constrained 
between minimum and maximum minihalo mass values $M^{\text{min}}_{\text{halo,0}}$ and $M^{\text{max}}_{\text{halo,0}}$ so that 
\begin{equation}
\frac{d\mathcal{P}_0(M_{\text{halo,0}})}{dM_{\text{halo,0}}} = C_{\text{halo,0}}\,  \text{exp}\left[ -\frac{(M_{\text{halo,0}}-\overline{M}_{\text{halo,0}})^2}{2\sigma_{\text{halo,0}}^2 %M_{\text{PBH}}^2
} \right] \text{    and    }    \int_{M^{\text{min}}_{\text{halo,0}}}^{M^{\text{max}}_{\text{halo,0}}} \frac{d\mathcal{P}(M_{\text{halo,0}})}{dM_{\text{halo,0}}} dM_{\text{halo,0}} = 1\,.\label{IMF} 
\end{equation}
The normalization constant $C_{\text{halo,0}}$ is given by 
\begin{equation}
C_{\text{halo,0}} = \frac{\sqrt{2/\pi}}{\sigma_{\text{halo,0}}
}\left[ \text{Erf}\left(  \frac{M_{\text{halo,0}}^{\text{max}}-\overline{M}_{\text{halo,0}}}{\sqrt{2}\sigma_{\text{halo,0}}%M_{\text{PBH}}
}\right) - \text{Erf}\left(\frac{M_{\text{halo,0}}^{\text{min}}-\overline{M}_{\text{halo,0}}}{\sqrt{2}\sigma_{\text{halo,0}}
}\right)  \right]^{-1}\approx 0.017 M_{\text{PBH}}^{-1}\,.    
\end{equation}
As a benchmark model, we take $\overline{M}_{\text{halo,0}} = 150 M_{\text{PBH}}$, $M_{\text{halo,0}}^{\text{min}} =  30 M_{\text{PBH}}$,
$M_{\text{halo,0}}^{\text{max}} =  270 M_{\text{PBH}}$
with $\sigma_{\text{halo,0}} = 24 M_{\text{PBH}}$ such that $M_{\text{halo,0}}^{\text{min}} \leq M_{\text{halo,0}} \leq M_{\text{halo,0}}^{\text{max}} $ correspond to $\pm 5 \sigma_{\text{halo,0}}$. 
In the expressions above, we have introduced the symbol $0$ to make it explicit that we refer to quantities before disruption effects. Later in the text we will suspend this for brevity. 

As galaxies evolve, further dressed PBHs and
axion DM background are incorporated to galactic halos.
After that dressed PBHs will undergo different levels of disruption depending on their orbital radii around the galactic center. Some of these dressed PBHs will undergo total disruption and some of them will undergo a partial mass loss so that the initial mass distribution given in Eq.~(\ref{IMF}) will be modified. We discuss 
%deal with 
the mass function after disruption in Sec.~\ref{IMFafterDisruption}.  

\section{Disruption of Dressed PBHs in the Milky Way}
After dressed PBHs are incorporated to the galactic halos at around $z\sim 6$~\cite{Cooray:2002dia},
their dark minihalos will undergo different levels of disruption depending on their minihalo masses, central PBH mass and radial orbit in the Galaxy. Generally speaking,
main sources of disruption for dark
matter substructures in the Milky Way are: (1) global tides coming from the
mean-field potential of the galaxy,
(2) high speed encounters with stars and
(3) tidal shocking during disk crossing.

Large dark matter substructures with masses about $10^7 M_{\odot}-10^9 M_{\odot}$ within a Galactocentric radius of $\sim30$ kpc
would undergo a  depletion of their abundance by a factor of (2-3) due to disk shocking~\cite{2010ApJ...709.1138D}. Axion miniclusters would undergo a tiny depletion of their abundances at the solar neighborhood, being the main source of disruption high speed encounters with stars. Taking a typical axion minicluster mass and radius of $\sim 10^{-12} M_{\odot}$ and $\sim 10^{7}\, \text{km}$, respectively, and assuming a power-law density profile ($\sim r^{-1.8}$), a depletion of about $(2-5)\%$ was reported in Refs.~\cite{Tinyakov:2015cgg, 2017JETP..125..434D}. Recently, a Montecarlo simulation was performed in Ref.~\cite{Kavanagh:2020gcy}, where an axion minicluster mass range of $\sim(10^{-19}-10^{-5})\,M_{\odot}$ is assumed. At the solar position, they reported survival probabilities of $99\%$ and $46\%$ for  power-law ($\sim r^{-9/4}$) and Navarro-Frenk-White~\cite{Navarro:1996gj} density profiles, respectively.

For the case of our interest, the cuspy density profile shown by dark minihalos around PBHs would offer them certain level of protection against disruption in the Milky Way. Analytical and numerical estimates performed by some of us in Ref.~\cite{Hertzberg:2019exb} show that typical dressed PBHs with $M_{\text{halo}}\simeq 100\,M_{\text{PBH}}$ lose about $9\%$ of their minihalo masses,
when circular orbits at the local neighborhood are assumed. Here, we extend our previous analysis performed in Ref.~\cite{Hertzberg:2019exb} to estimate the disruption map of dressed PBHs in the  Milky Way. We will focus on disruption coming from the mean field potential of the Milky Way and disk shocking when dressed PBHs cross the galactic disk. Disruption coming from high speed encounters with stars is disfavored for the PBH mass regime that interests us the most. As was discussed in Sec. 2.2 of Ref.~\cite{Hertzberg:2019exb}, the critical impact parameter $b_c$ at which the gained internal energy of the minihalo after the encounter, $\Delta E(b)$, is equal to the binding energy of the dressed PBH, $E_b$, reads as
\begin{equation}
b_c \sim 10^{-8} R_{\text{halo}} \left( \frac{M_{\star}}{M_{\odot}} \right)^{8/5} \left( \frac{220\,\text{km/s}}{V_{\text{rel}}} \right)^{8/5}     \left( \frac{M_{\odot}}{M_{\text{PBH}}} \right)^{16/15} \left( \frac{10^2\,M_{\text{PBH}}}{M_{\text{halo}}} \right)^{28/15}\,,
\end{equation}
where $V_{\text{rel}}$ is the relative velocity between both astrophysical objects and $M_{\star}$ is the star mass. Take as a Benchmark model $M_{\star} = M_{\odot}$ and $M_{\text{PBH}} = 30 M_{\odot}$. Only tiny impact parameters in terms of the minihalo radius are able to immediately disrupt the minihalo or lead to a significant minihalo mass loss. This kind of encounter is highly disfavored so that it is very unlikely that minihalos undergo one-off disruption. Thus, the total probability for disruption, $N_{\text{total}}$, is mostly given by  multiple star-dressed PBH encounters holding impact parameters such that $b > b_c$. The needed time for total disruption is longer than the Milky Way age in the parameter space of our interest even for large stellar densities, $n_{\star}$.  The probability for disruption from multiple encounters during a time $t$ is estimated as~\cite{Hertzberg:2019exb}
\begin{equation}
N_{\text{total}} \approx N_{\text{multiple}} = \frac{2\pi n_{\star} V_{\text{rel}} t}{E_b}\int_{b_c}^{\infty} \Delta E(b) b db.    
\end{equation}
Suppose that dressed PBHs undergo circular orbits close to the Galactic center at a radius of 100 pc. Thus, the needed time for minihalo disruption ($N_{\text{total}} = 1$) is estimated to be $t \gtrsim 200\,\text{Gyr}\, (21  \,\text{pc}^{-3}/n_{\star})$ for minihalos with masses $30\, M_{\text{PBH}} \lesssim M_{\text{halo}} \lesssim 270\, M_{\text{PBH}}$. We approximate the total stellar density at 100\, \text{pc} by the bulge contribution using the spherical version of the bulge profile described in Ref.~\cite{2017MNRAS.465...76M} (see below Eq.~(\ref{rhobulge}) and Table I in Sec.~\ref{MWMM}).

\subsection{Milky Way Mass Model}
\label{MWMM}

There is a rich literature  with respect to 
mass models of the Milky Way. The first example dates from the sixties in Ref.~\cite{1956BAN....13...15S}. Later, further 
models were put forward 
%examples were performed 
in Refs.~\cite{1998MNRAS.294..429D, 2002ApJ...573..597K, 2011MNRAS.414.2446M}. Here, we use the Milky Way model reported in Ref.~\cite{2017MNRAS.465...76M} (M16),  
which complies with theoretical modelling and fits the observational constraints. 
This mass model is composed of a spherically symmetric DM halo profile, a two-component
(thin/thick) galactic disk in cylindrical coordinates, a two-component ($H_I$ and molecular) gas disk and a
bulge profile.

The M16 model assumes a Navarro, Frenk, and White (NFW) profile~\cite{Navarro:1995iw} for the DM halo, which is parameterized as 
\begin{equation}
\rho_{\text{NFW}}(r) = \rho_s \left(  \frac{r}{r_s}\right)^{-1}\left(1+\frac{r}{r_s}\right)^{-2}\,,\\\label{rhoNFW}
\end{equation}
where $\rho_s$ and $r_s$ are the scale density and the scale radius, respectively, and $r$ is the Galactocentric radius.
This kind of profile is a common approximation to the density profile obtained in dark-matter-only cosmological simulations.

The two-component stellar disk is modelled in cylindrical coordinates $(z,r_{\text{cyl}})$ with an exponential decay as we departure from the Galactic plane. The density of both disk components 
%the stellar disks 
is modelled as
\begin{equation}
\rho_d(r_{\text{cyl}},z) = \frac{\Sigma_0}{2 z_d} \text{exp}\left( -\frac{|z|}{z_d} - \frac{r_{\text{cyl}}}{r_{\text{cyl,d}}}\right)\,,\label{rhodisk}
\end{equation}
where for each component  $z_d$ and $r_{\text{cyl,d}}$ are the scale height and scale length, respectively, and $\Sigma_0$ is the central surface density such that the total disk mass reads as $M_d = 2\pi\Sigma_0 r_{\text{cyl},d}^2$. In addition to the stellar disks, the M16 model includes a  two-component gas disk. This inclusion is necessary to make the number of stars with high vertical velocity in the local neighborhood dynamically consistent with the observed stars far from the Galactic plane.
The gas disk density profile reads as
\begin{equation}
    \rho_d(r_{\text{cyl}},z)=\frac{\Sigma_0}{4z_d}\text{exp}\left( -\frac{r_{\text{cyl,m}}}{r_{\text{cyl}}}-\frac{r_{\text{cyl}}}{r_{\text{cyl,d}}} \right) \text{sech}^2\left(\frac{z}{2z_d}\right)\,,
\end{equation}
where parameters $\Sigma_0, z_d, r_{\text{cyl, d}}$ are similar to those from the stellar disk case.
The parameter $r_{\text{cyl, m}}$ is the scale length associated with the central hole of the configuration. The density of the gas disks falls off as $\text{exp}(-z/z_d)$ as $z \rightarrow \infty$ in the same way as those for the stellar cases.

Lastly, the M16 model uses an axisymmetric bulge profile so that in cylindrical coordinates we have $r = \sqrt{r_{\text{cyl}}^2+(z/q)^2}$, where $q=0.5$ is the axial ratio.
We take for simplicity the spherical version of this profile so that the density profile reads 
 \begin{equation}
\rho_b(r) = \frac{\rho_{0,b}}{(1+r/r_0)^{\alpha_b}}\text{exp}\left[-\left( \frac{r}{r_{b}} \right)^2 \right] \,.\label{rhobulge}
\end{equation}

\begin{table*}
\begin{center}
\caption{Best-fit model for the Milky Way Galaxy performed in Ref.~\cite{2017MNRAS.465...76M} including the baryonic component parameters.}
\vspace{0.1cm}
\begin{tabular}{rrrrrr}
\hline
\rowcolor{Gray}
 $r_s [\textrm{kpc}]$& $\rho_s [\textrm{GeV}\textrm{cm}^{-3}]$&  $\alpha_b$&  $r_0 [\textrm{kpc}]$ & $r_b [\textrm{kpc}]$ & $\rho_{0,b} [M_{\odot} \textrm{pc}^{-3}]$\\   
\hline
19.6\textcolor{white}{99}&0.32\textcolor{white}{9999}&1.8&0.075\textcolor{white}{99}&2.1\textcolor{white}{99}&98.4\textcolor{white}{9999}\\
\hline
\end{tabular}
\end{center}
\end{table*}
\vspace{-0.5 cm}
\begin{table*}[hbt]
\begin{center}
\begin{tabular}{rrrr}
\hline
\rowcolor{Gray}
 $r_{\text{cyl,m}} [\textrm{kpc}]\hspace{1.1 cm}$& $r_{\text{cyl,d}} [\textrm{kpc}]\hspace{1.1 cm}$&  $z_d [\textrm{kpc}]\hspace{1.4 cm}$&  $\Sigma_0 [M_{\odot} \textrm{pc}^{-2}]\hspace{0.9 cm}$\\   
 \rowcolor{Gray}
       $(\textrm{thin}/\textrm{thick})\,(\textrm{H}_I/\textrm{H}_{II})$&  $(\textrm{thin}/\textrm{thick})\,(\textrm{H}_I/\textrm{H}_{II})$&  $(\textrm{thin}/\textrm{thick})\,(\textrm{H}_I/\textrm{H}_{II})$&  $(\textrm{thin}/\textrm{thick})\,(\textrm{H}_I/\textrm{H}_{II})$\\          
\hline
$(-/-)(4/12)\hspace{0.7 cm}$ &$(2.5/3.02)(7/1.5)\hspace{0.4 cm}$ & $(0.3/0.9)(0.085/0.045)$ & $(896/183)(53.1/2180)$\\
\hline
\end{tabular}
\end{center}
\end{table*}

Values of the parameters, including those associated with the baryonic components, for the best-fit model in Ref.~\cite{2017MNRAS.465...76M} (M16) are listed in Table I. For the bulge profile, we report the parameter values ($r_0, \alpha_b, r_b, \rho_{0,b}$) associated with the axysymmetric model but we approximate $q \approx 1$. 

\subsection{Global tides from the Milky Way}
\label{globaltides}

Unless dressed PBHs are orbiting 
close to the Galactic Center, the high density of dark minihalos around PBHs offers them protection against the tidal stripping. The overall Galactic potential acts on dressed PBHs stripping the region of the minihalo which is beyond the tidal radius, $r_{\text{tidal}}$. The stripped fraction of particles within the tidal radius is only a second-order correction corresponding to the highspeed
particle population close to the escape speed~\cite{Stref:2016uzb}.

Applying the distant-tide approximation, the tidal radius is calculated as~\cite{2008gady.book.....B}
\begin{equation}
r_{\textrm{tidal}} = \left(  \frac{M_{\textrm{halo}}(R_{\textrm{halo}})+M_{\textrm{PBH}}}{3 M_{\textrm{MW}}(R)} \right)^{1/3} \left( 1 - \frac{1}{3} \frac{d\textrm{ln}M_{\text{MW}}}{d\textrm{ln}R } \right)^{-1/3} R\,,\label{rtid}
\end{equation} 
where $R$ is the radial distance of the dressed PBH from the Galactic center under a circular orbit (or the perigalactic distance under eccentric orbits~\cite{1962AJ.....67..471K}) and $M_{\text{MW}}(R)$ is the mass of the Milky Way enclosed within the radius $R$ using the M16 model. The enclosed mass depend on the Galactic mass profile composed of the DM halo, the stellar and gas disks and the bulge. 

\begin{figure}[t!]
\centering
\includegraphics[width=11 cm]{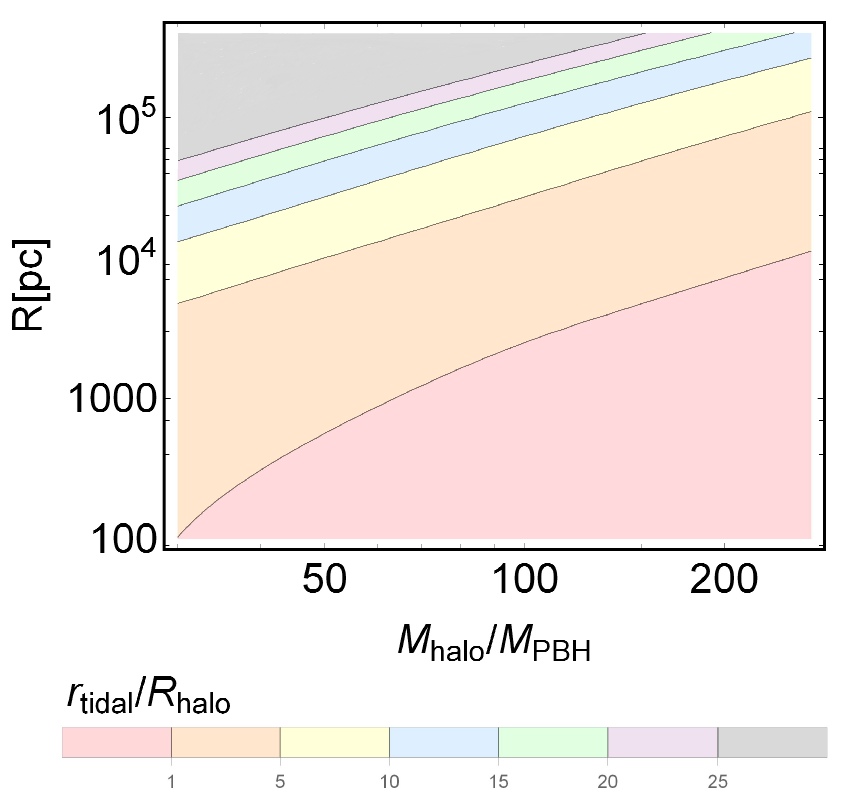}
\caption{Contour plot of the ratio $r_{\text{tidal}}/R_{\text{halo}}$ in the parameter space $(M_{\text{halo}}/M_{\text{PBH}},R)$ using Eq.~(\ref{rtid}).}
\label{Plot1}
\end{figure}

Using  $M_{\text{halo}} (R_{\text{halo}}) \gg M_{\text{PBH}}$ and the spherically symmetric mass profile of the minihalo, we have~\cite{Hertzberg:2019exb}
\begin{equation}
\hspace{-0.4cm}\frac{r_{\text{tidal}}}{R_{\text{halo}}} \sim 2 \left(\frac{R}{8.29\,\text{kpc}}\right) \left(\frac{10^2\,\text{M}_{\text{PBH}}}{M_{\text{halo}}}\right)  
\left( \frac{10^{11}\,M_{\odot}}{M_{\text{MW}}(R)} \right)^{1/3} \left( 1 - \frac{1}{3} \frac{d\textrm{ln}M_{\text{MW}}}{d\textrm{ln}R } \right)^{-1/3}\,, 
\end{equation} 
which depends on the minihalo mass in units of the central PBH mass rather than the $M_{\text{PBH}}$ itself. Figure~\ref{Plot1} shows a contour plot
for the ratio $r_{\text{tidal}}/R_{\text{halo}}$ in the parameter space $(M_{\text{halo}}/M_{\text{PBH}},R)$. The lighter the minihalo mass in units of the central PBH mass, the stronger its resistance against tidal stripping. For  $M_{\text{halo}} = 100\, M_{\text{PBH}} $ ( $M_{\text{halo}} = 30\, M_{\text{PBH}} $), we have $r_{\text{tidal}}/R_{\text{halo}} \gtrsim 1$
when dressed PBHs orbit around the Galactic Center at a closest radius $R \gtrsim 2.5\, \text{kpc}$
($R \gtrsim 0.1\, \text{kpc}$). %Based on this analysis, we redefine the radius of the minihalo as
%$\tilde{R}_{\text{halo}} = \text{min}(r_{\text{tidal}},R_{\text{halo}})$ and use this scale size to
%analyze tidal forces related to disk shocking in the next subsection.

\subsection{Gravitational Field of the Galactic disk}
\label{fsurv}

As dressed PBHs cross the Galactic plane, they undergo a compressive force coming from the gravitational field of the disk. This force pinches them along the normal to the disk plane so that the cumulative effect of successive disk crossings may eventually disrupt minihalos. This source of tidal disruption known as disk shocking was first proposed in Ref.~\cite{1972ApJ...176L..51O} in the context of globular clusters.

Modelling the Galactic disk as an infinite slab when dressed PBHs are just about to cross the disk plane and under the approximation of circular orbits for axion particles in minihalos, the gained energy of a minihalo per unit of axion mass in a single crossing
reads as (see Sec.~III.D.2 in Ref.~\cite{Stref:2016uzb} and Sec.~2.3 in Ref.~\cite{Hertzberg:2019exb})
\begin{equation}
\Delta E(r)  \approx \frac{32\pi^2 G_N^2  \rho^2_d(r_{\text{cyl}},0)\,z^2_d\,r^2}{3 V_z^2} A(\eta)\,,\label{egained}
\end{equation}     
where $\rho_d(r_{\text{cyl}},0)$ is the density profile of the thin disk at the Galactic plane and $A(\eta) = (1+\eta^2)^{-3/2}$ is the adiabatic correction which depends on the adiabatic parameter $\eta$. The adiabatic parameter is defined as $\eta(r,R) = \omega(r) \tau_{\text{cross}}(R)$, where $\omega(r)$ is the orbital frequency of the axion particles at distance $r$ from the central PBH and $\tau_{\text{cross}}(R)$ is the effective crossing time of dressed PBHs under circular orbits with a Galactocentric radius $R$. The impulse approximation holds when the axion orbital time within the minihalo is much longer than the disk crossing time, namely when $\eta(r) \rightarrow 0$ so that $A(\eta) \rightarrow 1$. By contrast, when  $\eta(r) > 1$ the efficiency of the disruption due to disk shocking starts to damp out. The angular momentum conservation associated with an axion particle orbiting several times during a single disk cross acts as a protection against disruption. 

We may estimate the axion orbital frequency by taking the inner dispersion velocity at radius $r$ as $\omega(r) = \sqrt{\langle  v_{\text{DM}}(r)^2\rangle} / r$. Using the isothermal approximation where the square of each Cartesian component of the velocity dispersion is equal to the half of the square of the circular velocity, we have  
\begin{equation}
\omega(r) = \sqrt{\frac{3 G_N m_{\text{halo}}(r)}{2 r^3}} \approx 0.2\,\textrm{Myr}^{-1}
 \left( \frac{10^2\,M_{\textrm{PBH}}}{M_{\text{halo}}} \right)^{3/2}   \left(\frac{R_{\text{halo}}}{r}\right)^{9/8}\,.\label{omega}
\end{equation}
The effective crossing time is estimated as function of the half-height $H$ of the disk and the vertical component of the dressed PBH velocity at a radius $R$ in the Galactic frame, $V_z(R)$. %Defining $M_{\text{MW}}(R)$ as the Milky Way mass within a 
%radius, 
We approximately have
\begin{equation}
\tau_{\text{cross}} = \frac{H}{V_z(R)} %= H \sqrt{\frac{2 R}{G_N M_{\text{MW}}(R)}}
\approx 0.67 \,\text{Myr} \left( \frac{H}{150\,\text{pc}} \right) \left(\frac{220\,\text{km/s}}{V_z}  \right)\,.  
\end{equation}
\begin{figure}[t!]
\centering
\includegraphics[width=11 cm]{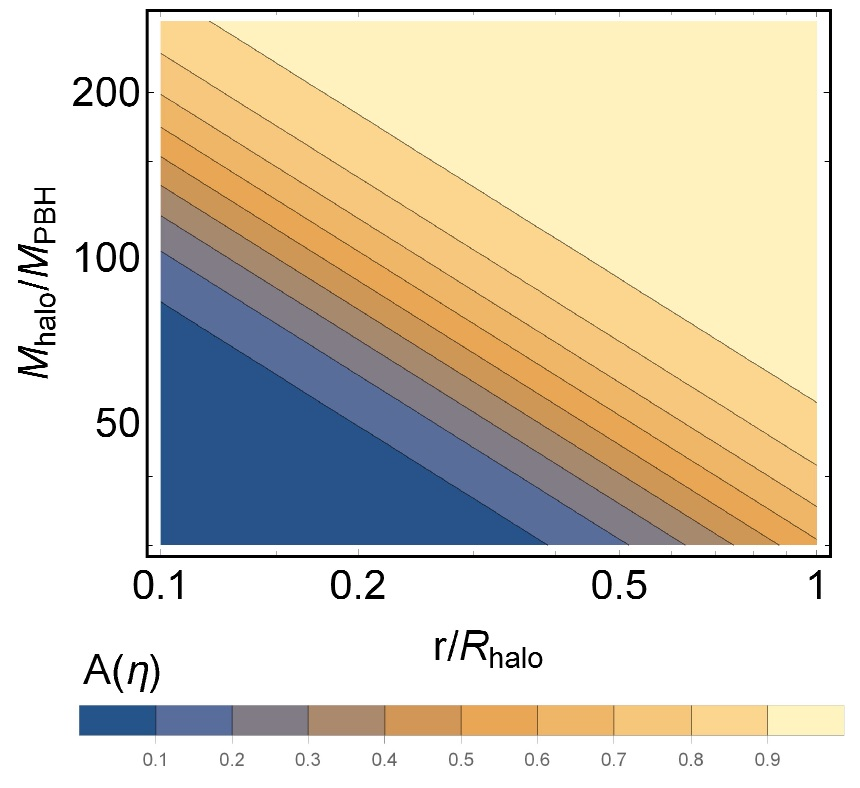}
\caption{Contour plot for the adiabatic correction $A(\eta)$ in the parameter space $(r/R_{\text{halo}}, M_{\text{halo}}/M_{\text{PBH}})$, where $r$ is the orbital radial of the axion particle within a minihalo of radius $R_{\text{halo}}$ and mass $M_{\text{halo}}$.}
\label{Plot2}
\end{figure}
Figure~\ref{Plot2} shows the contour plot for the adiabatic correction $A(\eta)$ considering a flat Galactic velocity curve of $\sim 220\, \text{km/s}$ 
in the parameter space of $(r/R_{\text{halo}},M_{\text{halo}}/M_{\text{PBH}})$. For heavier dressed PBHs, we have $A(\eta)\sim 1$, corresponding to the maximal efficiency of disk shocking, in most parts of minihalos apart from the 
very central part where the impulse approximation starts to break down. 
For lighter dressed PBHs the efficiency starts to damp out even 
in the outer parts of minihalos. For example, we see that for $150 \leq M_{\text{halo}}/M_{\text{PBH}} \leq 270$, we have $A(\eta) \sim (0.9-1)$ for $r/R_{\text{halo}} \gtrsim 0.3$, but for $40 \leq M_{\text{halo}}/M_{\text{PBH}} \leq 50$,  we have $A(\eta) \sim (0.8-0.9)$ for $r/R_{\text{halo}} \sim 1$. Thus, the lighter the dressed PBH the more resistant they are against disk shocking. Note that the adiabatic correction depends on the fraction $M_{\text{halo}}/M_{\text{PBH}}$ rather than  the central PBH mass as shown in Eq.~(\ref{omega}), when the axion orbital radius is expressed in terms of the minihalo radius. 

We assume dressed PBHs undergo circular orbits around the Galactic center during the age of the Milky Way, $T_{\text{MW}}\sim 10\,\text{Gy}$, so that the total number of disk crossing is calculated as
\begin{equation}
  N_{\text{cross}}(R) = T_{\text{MW}}\sqrt{\frac{G_N M_{\text{MW}}(R)}{\pi^2 R^3}}\,,
\end{equation}
where $M_{\text{MW}}(R)$ is the total Galactic mass within a radius $R$ using the M16 model. Before the first disk crossing, we take the minihalo radius to be  $R_{\text{halo}}(N_{\text{cross}}=0) = \text{min}(r_{\text{tidal}},R_{\text{halo}})$, where $r_{\text{tidal}}$ is the tidal radius considering global tides from the Milky Way as calculated in Sec.~\ref{globaltides}. The associated minihalo mass $M_{\text{halo}}(N_{\text{cross}}=0)$ is read from Eq.~(\ref{mr}). 

Define the gravitational potential of the minihalo at radius $r$ as
\begin{equation}
\phi_{\text{halo}}(r') = -G_N \int_{r'}^{\infty} dr\frac{m_{\text{halo}}(r)}{r^2}\,,
\end{equation}
so that after the first disk crossing we may recalculate the new tidal radius by comparing the gained energy of the minihalo per unit of axion mass, Eq.~(\ref{egained}), with the change in the gravitational potential as~\cite{Stref:2016uzb, Hertzberg:2019exb}
\begin{equation}
\Delta E (r_{\text{tidal}}) = -\left[ \phi_{\text{halo}}(r_{\text{tidal}})-\phi_{\text{halo}}(r') \right] = G_N \int_{r_{\text{tidal}}}^{r'} dr \frac{m_{\text{halo}}(r)}{r^2}\,,
\end{equation}
where $r'$ is the minihalo radius before the disk crossing.
The above equation is solved iteratively until the total number of disk crossings, $N_{\text{cross}}(R)$. The final tidal radius is related with the final tidal mass of the minihalo via Eq.~(\ref{mr}). 

Our analysis assume that between successive disk crossings dressed PBHs are able to virialize after a partial loss of particles. The virialization time is given by the free-fall time, $t_{\text{free-fall}} = 
\sqrt{3\pi/(32G_N \overline{\rho}_{\text{halo}})}$ where $\overline{\rho}_{\text{halo}} \sim 3M_{\text{halo}}/(4\pi R_{\text{halo}}^3)$. Since we are considering circular orbits around the Galactic Center with a Galactocentric radius $R$, we require that the half of the orbital period is at least the order of the free-fall time, so that
\begin{equation}
R \gtrsim R_{\text{halo}} \left( \frac{M_{\text{MW}}(R)}{8 M_{\text{halo}}} \right)^{1/3}\,.\label{virial}    
\end{equation}
For dressed PBHs with minihalo masses ranging as $30\, M_{\text{PBH}}\lesssim M_{\text{halo}} \lesssim 60 M_{\text{PBH}}$, this condition is satisfied until radii very close to the Galactic center, $R \gtrsim 100\, \text{pc}$.
However, as the minihalo mass increases the virialization assumption begins to break done for orbits with larger radii introducing a certain degree of inaccuracy in our results. For example, for a typical minihalo mass of $\text{M}_{\text{PBH}} \sim 100\, \text{M}_{\text{PBH}}$, we require $R \gtrsim 450\, \text{pc}$. Here we point out that the realistic picture includes eccentric orbits with much larger orbital periods relaxing Eq.~(\ref{virial}). 

\begin{figure}[t!]
\centering
\includegraphics[width=11 cm]{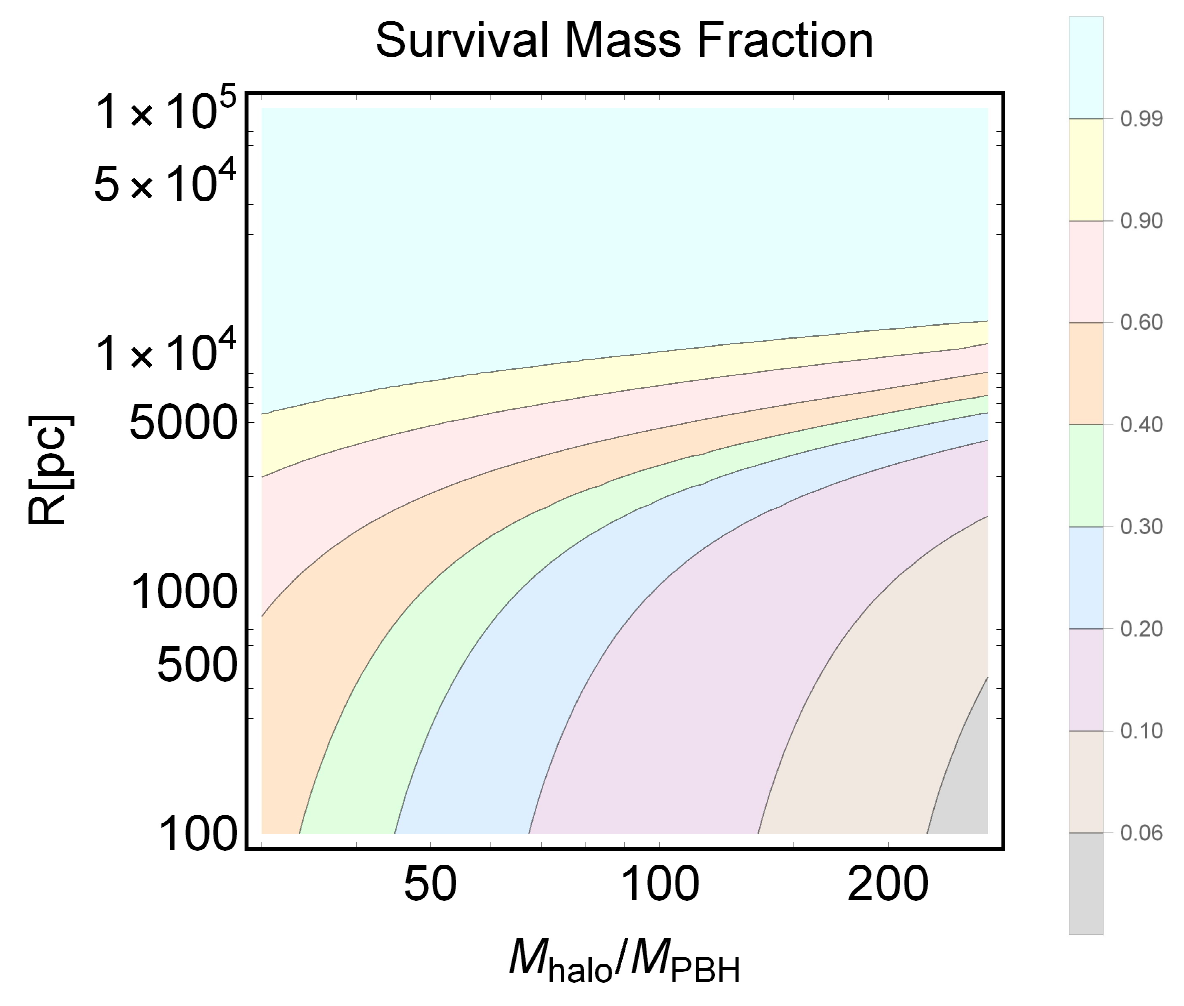}
\caption{Contour plot for the survival mass fraction of minihalos in the parameter space of $(M_{\text{halo}}/M_{\text{PBH}},R)$. Here we are assuming dressed PBHs undergo circular orbit around the Galactic center with a radius R[pc]. The minihalo mass sets in the horizontal axis refers to the minihalo before undergoing global tides forces and disk shocking. This initial mass is also used to calculate the survival mass fraction (see main text).}
\label{Plot3}
\end{figure}

Figure~\ref{Plot3} shows the contour plot for the survival mass fraction of minihalos in
the Milky Way in the parameter space  $(M_{\text{halo}}/M_{\text{PBH}},R)$, where we are assuming dressed PBHs undergo circular orbits of radius $R$ around the Galactic center. The survival mass fraction $f_{\text{surv}}$ is calculated by taking the ratio between the final minihalo mass after $N_{\text{cross}}$ crossings and the initial minihalo mass before undergoing global tides and disk shocking disruption.
In the whole range of minihalo masses studied by us, the mass loss is negligible when dressed PBHs undergo circular orbits around the Galactic center at a radius $R \gtrsim 13\, \text{kpc}$. As we approach the Galactic center, the mass loss begins to increase  and the survival mass fraction decreases down to 
$f_{\text{surv}} \sim (0.05-0.1)$ for $135\, \lesssim M_{\text{halo}}/M_{\text{PBH}} \lesssim 270\,$ at $ R \sim 100\, \text{pc}$. We see that the heavier the minihalo in units of the central PBH mass or the closer the minihalo orbital radius, the larger is the mass loss due to disruption in the Milky Way.

\subsection{Minihalo Mass Function after Disruption}
\label{IMFafterDisruption}
Since dressed PBHs undergo loss of mass due to tidal forces  and disk shocking in the Milky Way Galaxy,  
the disruption effects cause the mass function to deviate from its initial form Eq. (\ref{IMF}). 
Figure~\ref{Plot3} shows that there is a sizeable loss of mass from minihalos as the galactocentric radius decreases. This loss of mass will produce a  redistribution of the spectrum of masses for minihalos according to their orbital radius. If minihalos are not totally destroyed, which holds for the galactocentric radii considered in Fig.~\ref{Plot3}, the total number of minihalos is conserved so that we need to re-normalize the mass distribution after considering the effects of tidal forces from Milky Way.  
\begin{figure}[t!]
\centering
\includegraphics[width=11 cm]{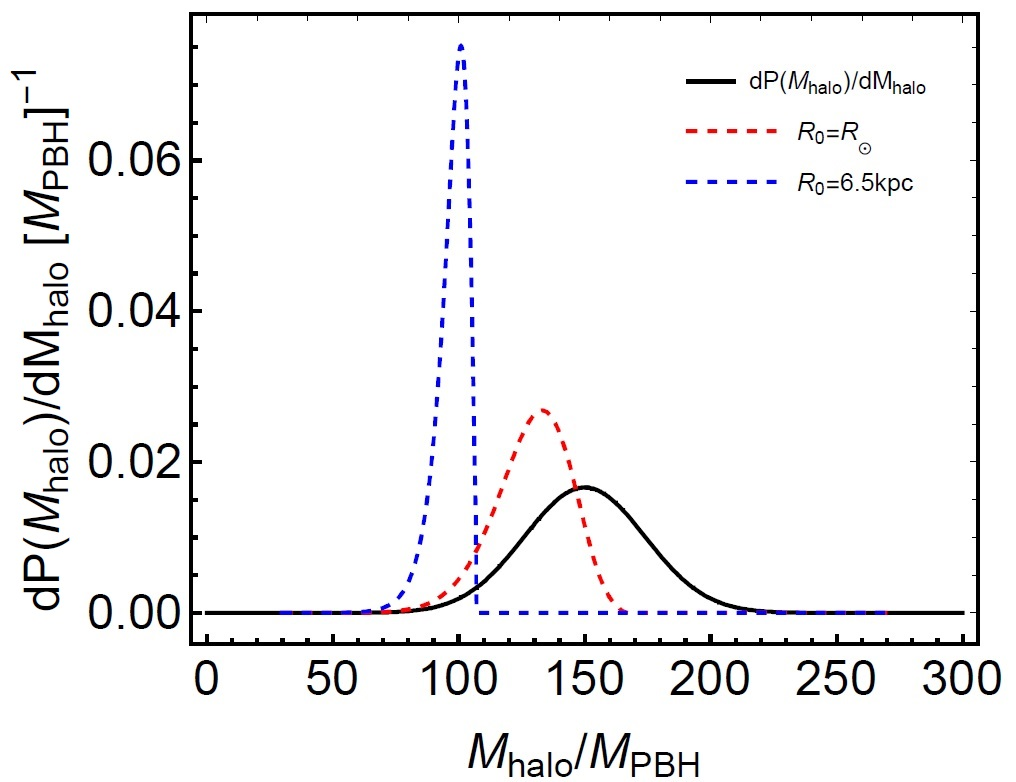}
\caption{Mass function $d\mathcal{P}/dM_{\text{halo}}$ at different galactocentric radii assuming circular orbits for dressed PBHs and the initial mass function (before disruption) given in Eq~(\ref{IMF}).}
\label{Plot4}
\end{figure} 
Figure~(\ref{Plot4}) shows the mass function after disruption. For a galactocentric radius $R \gtrsim 13\, \text{kpc}$ the 
disruption effects are negligible and the mass function practically coincides with the IMF
%effect on the IMF is negligible 
(solid black line). As the galactocentric radius becomes smaller,
the mass function begins to narrow and move towards lighter masses. At the solar position (red dashed line) and at 6.5 kpc (blue dashed line), most of minihalos lose about $\sim 11\%$ and $\sim 33\%$  of their initial masses, respectively. 
The mass spectrum redistribution leads to a significant deviation from the IMF which will further impact 
the encounter rate between NSs and dressed PBHs.   
\section{Encounter Rate for Neutron Stars-Dressed PBH encounters}
\label{sec:section2}
\subsection{Number Density for Neutron Stars}
There have been several efforts to model the NS distribution in the Milky Way (see e.g. Refs.~\cite{FaucherGiguere:2005ny, Bates:2013uma, Kaspi:2016jkv}). Neutron stars are mostly born during the core-collapse of massive stars. We may obtain a rough estimate of the total number of NSs produced in the Milky Way
through the present-day core-collapse supernovae 
rate~\cite{2006Natur.439...45D}. The total number of NSs born is about $10^8-10^9$, so that they represent a sizeable fraction of the Galactic stellar content. By performing Monte Carlo simulations to model NS orbits under suitable assumptions with respect to the Galactic potential, distribution of progenitors and birth velocities, Ref.~\cite{2010A&A...510A..23S}  reported that about 80 percent of NSs are in bound orbits and the remaining ones are unbound due to natal kicks. Following Ref.~\cite{2009PASP..121..814O}, where the space and velocity distributions of Galactic isolated old NSs is addressed, we estimate
that $40\%$ and $60\%$ of the total NS population were formed in the  Galactic disk and bulge, respectively. In particular, we take $\text{NS}_{\text{disk}} = 3.2\times10^8$ and $\text{NS}_{\text{bulge}}=4.8\times10^8$, so that $\text{NS}_{\text{total}}=\text{NS}_{\text{disk}}+\text{NS}_{\text{bulge}}=8\times10^8$. Here the interesting fact that the Galactic
bulge contains less stellar mass than the Galactic disk but more NSs are produced is because the bulge stellar mass function is skewed towards heavier stars than that in the disk~\cite{2009PASP..121..814O}. 

Due to the lack of pulsars observed in the inner regions
of the Milky Way, we do not have a direct measure of the isolated NSs which are distributed through the Galactic bulge.
While Ref.~\cite{Safdi:2018oeu} considered a Hernquist profile for that distribution, Ref.~\cite{Edwards:2020afl} assumed that this distribution tracks the stellar population in the bulge. We 
will assume that NSs tracks the stellar bulge so that the NS number density in the bulge $n_{\text{NS, b}}(r)$ is estimated as  
\begin{equation}
n_{\text{NS,b}}(r) = 0.038\, \frac{\rho_b(r)}{\overline{M}_{\text{NS}}}\,,\label{nNSbulge}     
\end{equation}
where $\overline{M}_{\text{NS}} = 1.4\,M_{\odot}$ is the characteristic mass for NSs having a characteristic radius of $R_{\text{NS}}=10\,\text{km}$ and the spherically symmetric stellar bulge density $\rho_b(r)$ is given by Eq.~(\ref{rhobulge}). The prefactor in Eq.~(\ref{nNSbulge}) comes from the integration
of the bulge stellar density to ensure a total number of NSs in the bulge equal to $\text{NS}_{\text{bulge}}$.

For the case of the NS distribution in the Galactic disk, we model the NS number density $n_{\text{NS},d}$ as a double exponential profile as follows
\begin{equation}
n_{\text{NS}, d}(r_{\text{cyl}}, z) = \frac{\text{NS}_{\text{disk}}}{4\pi r_{\text{NS}}^2 z_{\text{NS}}} \text{exp} \left( -\frac{|z|}{z_{\text{NS}} }-\frac{r_{\text{cyl}}^2}{2 r^2_{\text{NS}}} \right)\,,\label{nNSdisk}    
\end{equation} 
where $r_{\text{NS}} = 5\,\text{kpc}$ and $z_{\text{NS}} =1 \,\text{kpc}$ as was reported in Ref.~\cite{2010JCAP...01..005F} where the distribution of pulsars in the Galactic disk is studied in detail. 
\vspace{1 cm}
\subsection{Encounter Rate}

 We model the initial distribution of dressed PBHs per mass before any disruption effects as 
\begin{equation}
n_{\rm halo, 0}(r,M_{\rm halo,0}) = n_{\rm halo,0}(r)\frac{d\mathcal{P}_0(M_{\text{halo,0}},r)}{dM_{\text{halo,0}}}~,
\end{equation}
where the initial mass function  
$d\mathcal{P}_0(M_{\text{halo,0}},r)/dM_{\text{halo,0}}$ 
is given by Eq.~(\ref{IMF}).  From this we obtain the initial mass density of  dressed PBHs as $\rho_{\rm halo, 0}(r) = (\overline{M}_{\text{halo},0} + M_{\text{PBH}}) n_{\rm halo,0}(r)$, where $\overline{M}_{\text{halo},0}$ is the initial average minihalo mass for a PBH of mass $M_{\rm PBH}$. We assume the initial distribution $n_{\rm halo, 0}$ tracks the DM halo profile in the Galaxy and denote the inital fraction of DM in dressed PBHs before the disruption as $\overline{f}_{\text{halo,0}}\equiv \rho_{\rm halo,0}(r)/\rho_{\rm NFW}(r)$. The initial minihalo distribution can then be written as 
\begin{equation}
n_{\text{halo,0}}(r) = \frac{  \overline{f}_{\text{halo,0}}\, \rho_{\text{NFW}}(r)}{\overline{M}_{\text{halo,0}}+M_{\text{PBH}}} = \frac{f_{\text{PBH}}\,\rho_{\text{NFW}}(r)}{M_{\text{PBH}}}\,~,
\label{nr}
\end{equation}
where $f_{\rm PBH} = \rho_{\rm PBH}(r)/\rho_{\rm NFW}(r)$ and we have assumed $n_{\rm PBH} = n_{\rm halo,0}$, i.e. all the PBHs have accreted minihalos around them.

Including the disruption effects, the minihalo density per mass becomes 
\begin{equation}
n_{\rm halo}(r,M_{\rm halo}) = n_{\rm halo,0}(r)\frac{d\mathcal{P}(M_{\text{halo}},r)}{dM_{\text{halo}}}~,
\end{equation}
where $d\mathcal{P}(M_{\text{halo}},r)/dM_{\text{halo}}$ is the mass function with the disruption effects accounted for and it depends on $r$ as illustrated in Fig.~(\ref{Plot4}). Using this, we can estimate the  encounter rate between NSs and dressed PBHs as follows
\begin{equation}
    \Gamma_{\text{NS-PBH}} =  \int 
    \int \frac{dn_{\text{halo}}(r,M_{\rm halo})}{dR_{\text{halo}}} n_{\text{NS}}(r) \langle \sigma_{\text{eff}}(v_{\text{rel}})v_{\text{rel}} \rangle dR_{\text{halo}} d^3r\,,\label{GammaNSPBH}
\end{equation}
where $n_{\text{NS}}(r) = n_{\text{NS},d}(r_{\text{cyl}},z) + n_{\text{NS},b}(r)$ is the radial number density of neutron stars given by Eqs.~(\ref{nNSbulge}) and (\ref{nNSdisk}),  $\langle ... \rangle$ is the average over the NS-dressed PBH relative velocity distribution in the Milky Way halo, and $\sigma_{\text{eff}}$ is the effective cross section of the collision. Since $M_{\rm halo} = M_{\rm halo}(R_{\rm halo})$ we have 
 \begin{equation}
\left[ \frac{dn_{\text{halo}}(r,M_{\rm halo})}{dR_{\text{halo}}}\right] dR_{\text{halo}}   = \left[ n_{\text{halo,0}}(r) \frac{d\mathcal{P}(M_{\text{halo}},r)}{dM_{\text{halo}}}\right]dM_{\text{halo}}~.
 \end{equation}

 The effective cross section is given by the usual geometrical cross section plus the gravitational focusing enhancement as follows
 \begin{equation}
\sigma_{\text{eff}} (v_{\text{rel}}) = \pi (R_{\text{halo}} + R_{\text{NS}})^2\left[  1 + \left(\frac{v_{\text{esc}}^{\text{halo-NS}}}{v_{\text{rel}}}\right)^2\right] \simeq    \pi R_{\text{halo}}^2 \left(  1 + \frac{2G_NM_{\text{halo}}}{R_{\text{halo}}v_{\text{rel}}^2}\right)\,,\label{gf} 
 \end{equation}
 where $(v_{\text{esc}}^{\text{halo-NS}})^2 = 2G_N (M_{\text{halo}}+M_{\text{PBH}}+M_{\text{NS}})/(R_{\text{halo}}+R_{\text{NS}})$ is the mutual escape speed between the minihalo and the NS. We have taken $R_{\text{halo}} \gg R_{\text{NS}}$ and $M_{\text{halo}} \gg (M_{\text{PBH}},M_{\text{NS}})$ in the rightmost part of Eq.~(\ref{gf}). The gravitational enhancement, which comes from
the deflection of trajectories by the gravitational
attraction between the NS and the minihalo, can be written as
\begin{equation}
\frac{2G_N M_{\text{halo}}}{R_{\text{halo}}v_{\text{rel}}^2}\sim 10^{-6} \left( \frac{150\,M_{\text{PBH}}}{M_{\text{halo}}} \right)^{1/3} \left( \frac{M_{\text{PBH}}}{M_{\odot}} \right)^{2/3} \left( \frac{2\times220\,\text{km/s}}{v_{\text{rel}}} \right)^2 \,.\label{gravfoc} 
\end{equation}
 Suppose that dressed PBHs and NSs have velocities dispersion $\sigma_{\text{PBH}}$ and  $\sigma_{\text{NS}}$, respectively, and both follow a Maxwell-Boltzmann distribution so that the probability $P(v_{\text{rel}})dv_{\text{rel}}$ that dressed PBHs and NSs have a relative speed in the interval $(v_{\text{rel}}+dv_{\text{rel}})$ reads as (see page 712 in Ref.~\cite{2008gady.book.....B})
 \begin{equation}
 P(v_{\text{rel}})dv_{\text{rel}} = \frac{C_0}{\left(2\pi \sigma_{\text{rel}}^2\right)^{3/2}}\text{exp}\left(- \frac{v^2_{\text{rel}}}{2\sigma_{\text{rel}}^2} \right)v_{\text{rel}}^2dv_{\text{rel}}\,,\label{MB}    
 \end{equation}
where $\sigma_{\text{rel}} = \sqrt{\sigma_{\text{PBH}}^2 + \sigma_{\text{NS}}^2} \approx  \sqrt{2}\times\sqrt{G_N M_{\text{MW}}(R)/R} \approx \sqrt{2}\times220\,\text{km/s}$ is the relative velocity dispersion assuming a flat Galactic velocity curve and $C_0$ is the normalization constant so that $4\pi \int_0^{v_{\text{esc}}}  P(v_{\text{rel}})dv_{\text{rel}} = 1$, with $v_{\text{esc}}$ as a characteristic escape velocity in the Galaxy. In most cases of interest, we have $v_{\text{esc}} \gtrsim \sqrt{2} \sigma_{\text{rel}}$ leading to $C_0\approx 1$ (e.g., see Fig. 3 in Ref.~\cite{2018A&A...616L...9M}). Thus, we may estimate the average over the relative velocity of the effective cross section as
\begin{equation}
\langle \sigma_{\text{eff}}(v_{\text{rel}})v_{\text{rel}}\rangle = 4\pi \int_0^{v_{\text{esc}}} \sigma_{\text{eff}}(v_{\text{rel}}) v_{\text{rel}} P(v_{\text{rel}})dv_{\text{rel}}\approx \sqrt{8\pi \sigma_{\text{rel}}^2} R_{\text{halo}}^2\left( 1 + \frac{G_N M_{\text{halo}}}{R_{\text{halo}}\sigma_{\text{rel}}^2}   \right)\,,\label{sigmav}    
\end{equation}
where the approximation is obtained by pushing the upper limit of the integral to infinity supported by the quick decay of the exponential. Considering that   $R_{\text{halo}} = R_{\text{halo}}(M_{\text{halo}})$ via Eq.~(\ref{mr}), we put all together so that the encounter rate in the Milky Way in Eq.~(\ref{nr}) reads  
\begin{equation}
\Gamma_{\text{NS-PBH}}\approx \sqrt{8 \pi \sigma^2_{\text{rel}}} \int \int  
n_{\text{halo,0}}(r) \frac{d\mathcal{P}(M_{\text{halo}},r)}{dM_{\text{halo}}}n_{\text{NS}}(r) R_{\text{halo}}^2\left(1+\frac{G_N M_{\text{halo}}}{R_{\text{halo}}\sigma^2_{\text{rel}}}  \right) dM_{\text{halo}}d^3r\,.\label{erate}
\end{equation}
As shown before, the tidal disruption due to the Milky Way mean field potential and disk shocking both depend on the ratio $M_{\text{halo}}/M_{\text{PBH}}$ rather than $M_{\text{PBH}}$. Thus the survival mass fraction in Fig.~\ref{Plot3} does not depend directly on the PBH mass. However, the encounter rate shows a direct dependence on $M_{\text{PBH}}$ through the minihalo radius and the radial number density of dressed PBHs. From Eq.~(\ref{gravfoc}), we see that the gravitational focusing may be safely neglected for $M_{\text{PBH}} \sim(1-10)M_{\odot}$. On the other hand, $R_{\text{halo}} \propto (M_{\text{PBH}}/M_{\odot})^{1/3}$ and $n_{\text{halo}}(r) \propto (M_{\text{PBH}}/M_{\odot})^{-1}$ so that the differential encounter rate runs with the PBH mass as $d\Gamma_{\text{NS-PBH}}/dr \sim (M_{\text{PBH}}/M_{\odot})^{-1/3}$. The larger the central PBH mass, the smaller the encounter rate due to the decrease of the radial number density of dressed PBHs which dominates over the increase of the cross section.
 
\begin{figure}[t!]
\centering
\includegraphics[width=11 cm]{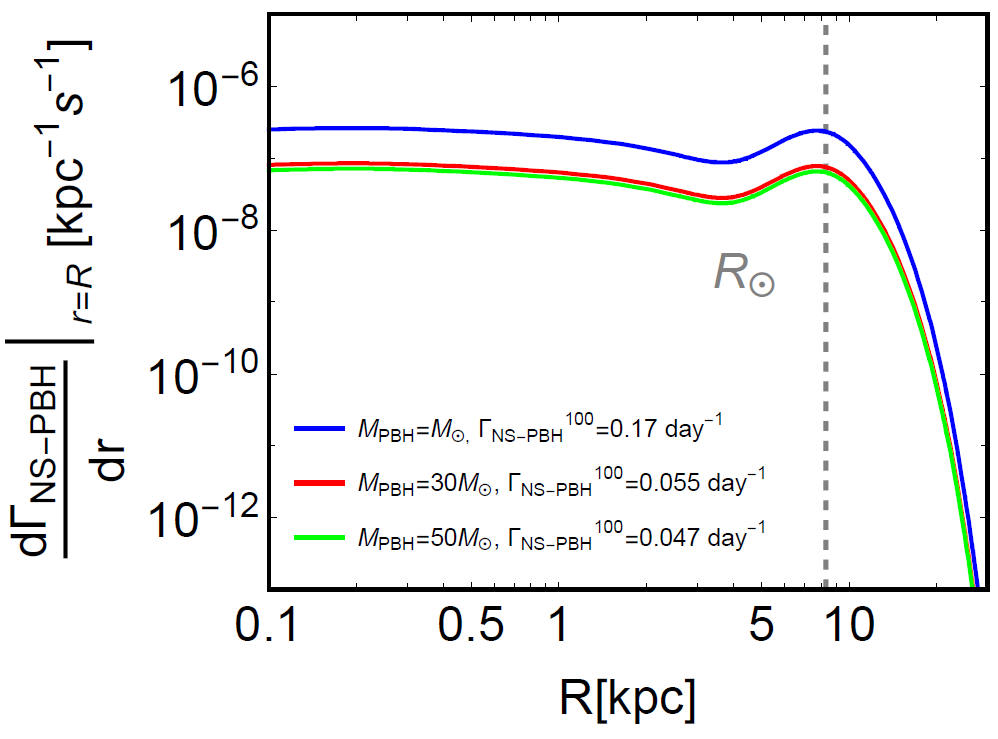}
\caption{Differential encounter rate between NSs and dressed PBHs, $d\Gamma_{\text{NS-PBH}}/dr$, in terms of the galactocentric radius $R\, \left[\text{kpc}\right]$ according to Eq.~(\ref{erate}). We  assume a Gaussian distribution for the initial mass function of dressed PBHs (before disruption) and take into account the minihalo mass loss due to global tides and disk shocking. We assume dressed PBHs undergo circular orbits around the Milky Way center. The fraction of DM in PBHs is fixed to  $f_{\text{PBH}}=10^{-3}$ and the central PBH mass is taken to be $M_{\text{PBH}}=(1,30,50)M_{\odot}$ for the blue, red, and green solid lines, respectively. The total encounter rate integrated up to $R \geq 100\, \text{pc}$ and referred to as $\Gamma_{\text{NS-PBH}}^{100}$ is indicated in the plot for each case. The vertical dashed gray line indicates the solar position.}
\label{Plot5}
\end{figure}

Figure~\ref{Plot5} shows the differential encounter rate between NSs and dressed PBHs with respect to the galactocentric radius assuming a Gaussian distribution for the initial mass function of dressed PBHs in Eq.~(\ref{IMF}) and taking into account the mass loss of minihalos due to tidal forces and disk shocking in the Milky Way. We have assumed that dressed PBHs undergo circular orbits around the Milky Way center.  The fraction of DM in PBHs is taken to be $f_{\text{PBH}}=10^{-3}$. As shown in Fig.~\ref{Plot3}, tidal forces acting on dressed PBHs have negligible effects for $R\gtrsim 13\, \text{kpc}$ so that the quick drop of the differential encounter rate for larger galactocentric radii in Fig.~\ref{Plot5} is due to the decrease in the numbers of NSs and dressed PBHs available for collisions. As we approach the innermost parts of the Milky Way there are two factors which compete against each other. From one side the number of NSs and dressed PBHs increases but on the other hand the mass distribution of dressed PBHs moves towards
lighter minihalo masses as shown in Fig.~\ref{Plot4}. Both effects tend to compensate each other at distances of $\mathcal{O}(100) \text{ pc}$ leading to an  encounter rate of $\mathcal{O}(10^{-7})\,\text{kpc}^{-1}\text{s}^{-1}$.  As explained above, the heavier the central PBH mass, the smaller the differential encounter rate at the same galactocentric radii. The blue, red and green solid lines correspond to the cases $M_{\text{PBH}}=(1,30,50)M_{\odot}$, respectively, showing  
total encounter rates for $R \gtrsim 100\, \text{pc}$ as $\Gamma^{100}_{\text{NS-PBH}} \approx (0.17, 0.055, 0.047)\, \text{day}^{-1}$, respectively. These encounter rates are about two orders lower than the rates reported in Ref.~\cite{Edwards:2020afl} for the case of axion miniclusters colliding with NSs, when these dark matter substructures are assumed to have a power-law density profile, $\sim r^{-9/4}$, and constitute the entire  DM.  

\section{Radio Signal via QCD-axion resonance and Detection}
\label{sec:section3}
\subsection{Transient Radio Signal}

As the NS goes through the minihalo of a dressed PBH, axion particles fall
towards the NS surface and they resonantly convert into photons within its magnetosphere.
For the axion mass in the MHz-GHz range, the conversion takes place in a small region around $r_c > R_{\text{NS}}$, where $r_c$ is called the conversion radius~\cite{Hook:2018iia}. This particular radius corresponds to the place at which the plasma frequency $\omega_p$ matches the axion mass $m_a$. For simplicity and for the sake of comparison with Ref.~\cite{Edwards:2020afl} where aligned NSs are considered, we focus on the  case of aligned or slightly oblique NSs by assuming that the misalignment angle between the magnetic field and the NS rotating axis is small, $\theta_m \ll 1$. 
%case in which the magnetic field shows a small misalignment with the NS rotating axis, so that the angle between them is $\theta_m \ll 1$ leading to a slightly oblique NSs. In such scenario, 
In this case, the magnetosphere can be reasonably approximated by the Goldreich-Julian model~\cite{1969ApJ...157..869G} and the power radiated per unit solid angle at the observation angle $\theta$ and to zeroth order in $\theta_m$ in the WKB and stationary phase approximations reads as~\cite{Hook:2018iia}
\begin{equation}
\frac{d\mathcal{P}(\theta)}{d\Omega} \approx \frac{ g_{a\gamma\gamma}^2 B_0^2 \rho_a(r_c) \pi v_c }{6 m_a} \left(\frac{R_{\text{Ns}}^2}{r_c}\right)^3 \left[3\text{cos}^2(\theta)+1\right]\,,\label{DPDOMEGA}  
\end{equation}
where the conversion radius is given by  
\begin{equation}
 r_c(\theta) = 224\, \text{km} \left( \frac{R_{\text{NS}}}{10\,\text{km}}\right) \left[\left(\frac{B_0}{10^{14}\,\text{G}} \right) \left(\frac{1\,\text{sec}}{P} \right) \left(\frac{1\,\text{GHz}}{m_a} \right)^{2}\right]^{1/3}|3\text{cos}^2(\theta)-1|^{1/3}\,.\label{rconv} 
\end{equation}
Here $B_0$ is the magnetic field strength at the NS poles, $\rho_a (r_c)$ and $v_c$ are the axion density and velocity at a distance $r_c$ from the NS, respectively, and $P$ is the NS spin period.

\begin{figure}[t!]
\centering
\includegraphics[width=15.7 cm]{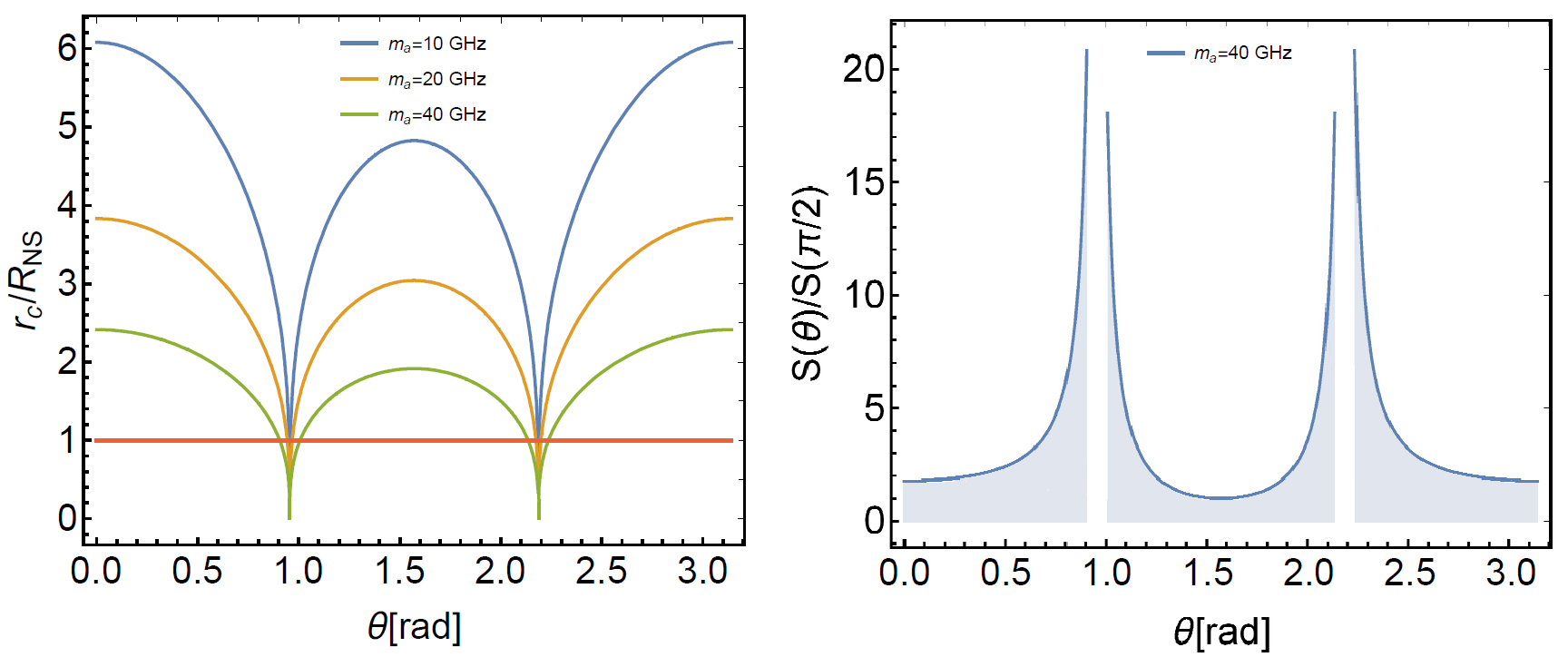}
\caption{(Left) Ratio between the conversion radius and the NS radius using Eq.~(\ref{rconv}) for different axion masses, $m_a = (10,20,40)\,\text{GHz}$, in function of the polar angle $\theta$. We have fixed all remaining parameters, $B_0 = 10^{14}\,\text{G}, R_{\text{NS}}=10\,\text{ km}, P = 1\,\text{s}, R_{\text{NS}}=10\,\text{km}$. The horizontal red line indicates when $r_c(\theta)/R_{\text{NS}} = 1$. (Right) The spectral flux density at given polar angle normalized by the spectral flux density at $\theta=\pi/2$ in function of the polar angle. The shaded region indicates the parameter space at which $r_c(\theta) > R_{\text{NS}}$.  }
\label{Plot5bb}
\end{figure}

\begin{figure}[t!]
\centering
\includegraphics[width=15.5 cm]{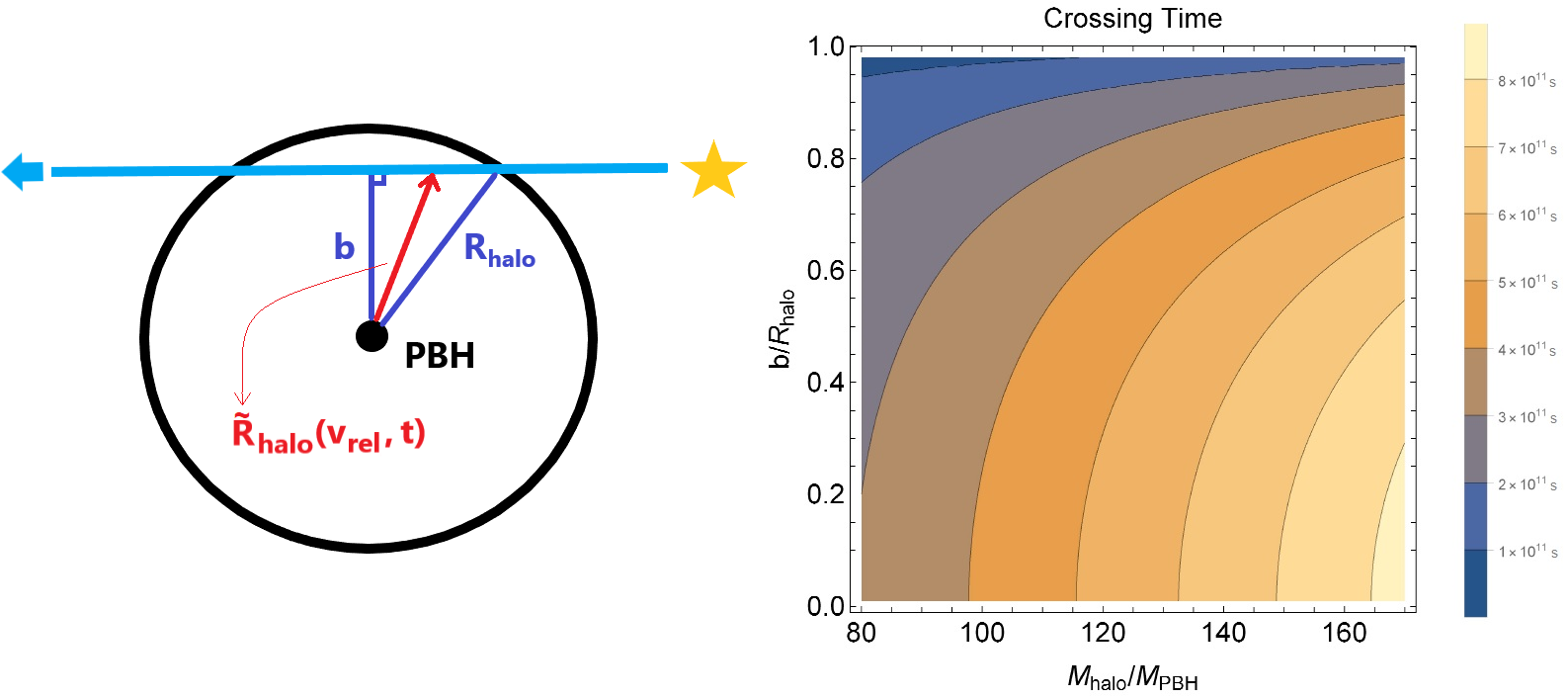}
\caption{(Left) Schematic picture of a NS-dressed PBH encounter illustrating the quantities defined in the text.
(Right) Contour plot for the crossing time $T_{\text{cross}}$ in the parameter space of $(M_{\text{halo}}/M_{\text{PBH}}, b/R_{\text{halo}})$. We have considered the spectrum of expected minihalo masses (after disruption) located at the local neighborhood and $v_{\text{rel}} = \sqrt{2}\sigma_{\text{rel}}\approx 2\times 220\,\text{km/s}$.}
\label{Plot5b}
\end{figure}

Now we consider the individual trajectories of the axion particles close to the NS surface. The maximum impact parameter $b_{\text{a, max}}$, which ensures that an axion particle ends up crossing the conversion radius, is given by 
\begin{align}
&b_{\text{a, max}}(r_c) \approx  \, r_c \left( 1 + \frac{2G_N M_{\text{NS}}}{r_c v_{\text{rel}}^2} \right)^{1/2}\,,\\
\sim & \,7 \times 10^{-12}\,\text{pc} \,\left(\frac{r_{\rm c}}{224 {\rm km}}\right)\left[ 1 + 6\times 10^4 \left( \frac{224\,\text{km}}{r_c}\right) \left( \frac{2\times\,220 \text{km/s}}{v_{\text{rel}}} \right)^2 \left( \frac{M_{\text{NS}}}{M_{\odot}} \right)  \right]^{1/2},
\end{align}
which is associated with an axion velocity at $r_c$ given by 
\begin{equation}
  v_c \approx \left(\frac{2G_N M_{\text{NS}}}{r_c}\right)^{1/2} \sim 0.1 \text{c}  \left(\frac{M_{\text{NS}}}{M_{\odot}}  \right)^{1/2} \left( \frac{224\,\text{ km}}{r_c} \right)^{1/2}\,.  
\end{equation}
At distances of the order of $b_{\text{a, max}}$, the axion particle mostly feels the NS gravitational potential and the gravitational potential from the effective mass located at the center of the dressed PBH may be safely neglected. In addition, we have neglected velocity dispersion
%the inner dispersion velocity 
of the axion particles within the minihalo because, using the isothermal approximation, we have 
\begin{align}
\sqrt{ \langle v_{\text{DM}}(r)^2 \rangle } &= \left (  \frac{3G_N m_{\text{halo}}(r)}{2r} \right)^{1/2}\,,\\ 
&\simeq 0.8\,\text{km/s}\,\left( \frac{10^2\,M_{\text{PBH}}}{M_{\text{halo}}} \right)^{1/6} \left( \frac{M_{\text{PBH}}}{M_{\odot}} \right)^{1/3} \left(\frac{10^{-2}\,R_{\text{halo}}}{r}\right)^{1/8}\,.\label{vdm}
\end{align} The above analysis only breaks down if the impact parameter between the NS and the center of mass of the dressed PBH is extremely small but such encounters are statistically disfavored.  Suppose that the NS crosses the minihalo with an impact paramater $b$ as indicated in Fig.~\ref{Plot5b} (left). The crossing time is estimated to be $T_{\text{cross}} = (2/v_{\text{rel}})  \sqrt{R_{\text{halo}}^2-b^2}$. Substituting the 
expressions for $v_c$ and $r_c$ into Eq.~(\ref{DPDOMEGA}), we obtain 
\begin{equation}
\frac{d\mathcal{P}(\theta)}{d\Omega} = \frac{d\mathcal{P}(\theta=\pi/2)}{d\Omega} \times \frac{(3\text{cos}^2(\theta)+1)}{|3\text{cos}^2(\theta)-1|^{7/6}}\,,\label{split}  
\end{equation}
where 
\begin{align}
\frac{d\mathcal{P}(\theta=\pi/2)}{d\Omega} \sim 7\times 10^7&\, \text{W} \left( \frac{R_{\text{halo}}}{\widetilde{R}_{\text{halo}}} \right)^{9/4} \left( \frac{10^2 M_{\text{PBH}}}{M_{\text{halo}}} \right)^{3} \left( \frac{g_{a\gamma\gamma}}{10^{-12}\,\text{GeV}^{-1}}\right)^{2} \left( \frac{R_{\text{NS}}}{10\,\text{km}} \right)^{5/2} \,\nonumber\\
&\times \left( \frac{m_a}{\text{GHz}} \right)^{4/3}\left( \frac{B_0}{10^{14}\,\text{G}} \right)^{5/6}\left(  \frac{P}{1\,\text{s}}\right)^{7/6} \left(  \frac{M_{\text{NS}}}{M_{\odot}}\right)^{1/2}\,.\label{DPDOmeganoangle}
\end{align}
Here we have taken $\rho_a(r_c) \approx \rho_{\text{halo}}(\widetilde{R}_{\text{halo}})$ given by Eq.~(\ref{rho}) for $r_c \ll \widetilde{R}_{\text{halo}} \lesssim R_{\text{halo}}$ and $\widetilde{R}_{\text{halo}}^2= b^2 + \left(\sqrt{R_{\text{halo}}^2-b^2}-v_{\text{rel}}t\right)^2$. 
Equation~(\ref{split}) is only valid for polar angles at which $r_c(\theta) > R_{\text{NS}}$. From Eq.~(\ref{rconv}), we see that the set of angles,  $\{\Theta\}$,  which satisfy this condition correspond to
\begin{equation}
\Theta = [0, \theta_1[\, \cup \,]\theta_2, \theta_3[\, \cup \,]\theta_4, \pi]\, 
\end{equation}
where $0 \leq \theta_1 < \theta_2 < \theta_3 < \theta_4 \leq \pi$ and
\begin{align}
\theta_{1} &= \text{Cos}^{-1}\left[ \sqrt{\frac{1}{3} \bigg(
1 + \frac{1}{22.4^3}\left(\frac{m_a}{1\,\text{GHz}}\right)^2\left(\frac{P}{1\,\text{s}}\right)\left(\frac{10^{14}\,\text{G}}{B_0}\right)\bigg )} \right]\,\hspace{0.3 cm}\,\text{and}\hspace{0.3 cm} \theta_4 &= \pi - \theta_1\,\label{theta1} \\
\theta_{2} &= \text{Cos}^{-1}\left[ \sqrt{\frac{1}{3} \bigg( 1 - \frac{1}{22.4^3}\left(\frac{m_a}{1\,\text{GHz}}\right)^2\left(\frac{P}{1\,\text{s}}\right)\left(\frac{10^{14}\,\text{G}}{B_0}\right)\bigg )} \right]\,\hspace{0.3 cm}\,\text{and}\hspace{0.3 cm} \theta_3 &= \pi - \theta_2\,\label{theta3}
\end{align}
The allowed angles depend on the axion mass and the NS spin period and the magnetic field at poles. Figure~\ref{Plot5bb} (left) shows the ratio $r_c / R_{\text{NS}}$ in terms of the polar angle for several axion masses with all other parameters fixed. As the axion mass increases, 
the angular regions corresponding to the absence of resonant axion-photon conversion slowly begin to grow until they eventually extend over the full angular space, i.e. $\Theta = \{ \}$.

We take the axion mass as a free parameter
within the 
%suitable 
range $m_a \in [m_{a,\text{lower}},m_{a,\text{upper}}]$   constrained as follows. The 
 lower limit
$m_{a,\text{lower}}$ is determined by the upper bound on the axion decay constant, so that $m_{a,\text{lower}}(F_a = 10^{12}\,\text{GeV}) \simeq \,\text{GHz}$. The 
 upper limit $m_{a,\text{upper}}$
is set either by (i) the maximum mass for which the conversion radius is larger than the NS radius, Eq.~(\ref{rconv}), or (ii) the maximum frequency detectable by a given radio telescope, depending on which one gives a stronger bound.

\textcolor{black}{The photon flux is peaked around the central frequency $\nu_{\text{peak}} = m_a/(2\pi)$. Assuming energy conservation, the signal bandwidth is taken to be proportional to the initial DM dispersion as $\Delta \nu \sim \nu_{\text{peak}} \langle v_{\text{DM}}^2\rangle$ in Ref.~\cite{Hook:2018iia}. For the case of axion DM background, this bandwidth is typically of the order of $\text{kHz}$, but in our setup where the initial DM dispersion is set by Eq.~(\ref{vdm}), this would lead to a narrower spectral line. In Ref.~\cite{Edwards:2020afl}, where the resonant signal is studied in the context of NS-axion minicluster encounters, the signal bandwidth is also predicted to be very narrow
due to the small internal velocity dispersion of these astrophysical objects.  These authors 
 argue that the signal bandwidth is expected to be widened
%assume that probably the signal bandwidth is widened 
by other mechanisms and fix the bandwidth of the signal to 1 kHz,  which equals
%according to 
the resolution of current and prospective radio
telescopes~\cite{SKAGuide,  2011ApJ...739L...1P, 2019arXiv191212699B}. In the context of oblique NSs and when axions are non-relativistic at the conversion surface, authors in Ref.~\cite{Battye:2019aco}  point out that the broadening of the spectral line should be dominated by the relative motion between the resonant surface and the observer instead of the initial axion DM dispersion.
In this case, the bandwidth of the signal would read as (see Eq.~(84) in Ref.~\cite{Battye:2019aco})
\begin{equation}
\Delta \nu = 7\,\text{MHz} \left( \frac{2\pi\, \text{s}}{P} \right)^{4/3} \left( \frac{m_a}{6.6\times 10^{-6}\,\text{eV}} \right)^{1/3} \left( \frac{B_0}{10^{14}\,\text{G}} \right)^{1/3}\epsilon^2\,, \label{B}
\end{equation}
where $\epsilon$ is a geometrical factor which depends of the NS properties. For an oblique NS, the intersection of the conversion surface with a plane perpendicular to the NS rotation axis forms an ellipse with eccentricity $\epsilon$ rather than a circle. As an example, consider the particular case of the isolated NS RX J0806.4-4123 located at about $250\,\text{pc}$ distance from the Earth, which has a period $P\approx 11.37\,\text{s}$ and a magnetic field at poles $B_0 \approx 2.5\times10^{13}\,\text{G}$~\cite{Kaplan:2009ce}.  For axion masses in the range $300\, \text{MHz} \lesssim m_a \lesssim 50\, \text{GHz}$, the signal bandwidth  $\Delta \nu\approx \text{kHz}$ for $\epsilon \simeq (0.01-0.03)$ and $ \Delta\nu \approx 100\, \text{kHz}$ for $\epsilon \simeq (0.1-0.3)$. For the sake of comparison with Ref.~\cite{Edwards:2020afl}, we will first  set the signal bandwidth equal to 1 kHz and analyze the parameter space for detection in this case. In \cite{Edwards:2020afl} the value 1 kHz was taken to estimate the bandwidth broadening due to astrophysical effects not specified to further detail. We follow the same phenomenological approach here but point out in addition that requiring the Doppler broadening effect (\ref{B}) to be below 1 kHz level requires small eccentricity, $\epsilon \sim 10^{-3}$ for the parameters shown in our figures.
%assume a Doppler broadening leading to a signal bandwidth equal to 1 kHz and analyze the parameter space for detection. 
Later, we will also consider larger bandwidths  dominated by the Doppler effect  Eq.~(\ref{B}) to compare our results with those in Ref.~\cite{Battye:2019aco}.}

The spectral flux density is calculated as $S(\theta) = (d\mathcal{P}(\theta)/d\Omega)/(d^2 \Delta \nu)$, where $d$ is the distance from the encounter to the Earth and $\Delta \nu$ is signal bandwidth as explained above. Due to the physical extension of the minihalo, we have a transient radio signal as NS crosses it. We will present our results in terms of 
the mean spectral density $\langle S(\theta) \rangle = (1/\Delta t) \int S(\theta) dt$, where the time lapse is given by $\Delta t$. From Eq.~(\ref{split}), we see that the spectral flux has a strong dependence on the polar angle, so that $S(\theta)  \in [S_{\theta,\text{min}}, S_{\theta,\text{max}} ]$, where $S_{\theta,\text{min}} \equiv S(\theta=\pi/2)$ and $S_{\theta,\text{max}} \equiv  \text{lim}_{\theta \rightarrow \theta_1} S(\theta)=\text{lim}_{\theta \rightarrow \theta_4} S(\theta)$.
%S(\theta = \text{lim}_{\theta \rightarrow \theta_1} \theta) =  S(\theta = \text{lim}_{\theta \rightarrow \theta_4} \theta)$}. 
The signal is very sharply peaked around the maxima, as demonstrated in Fig.~\ref{Plot5bb} (right) for a given parameter set, and tracking a maximum for an extended time period with a sufficiently high angular precision may be technically challenging. For the sake of completeness, we will present our results using different angles including $\theta = \pi/2$.  

As the NS crosses the dressed PBH, the axion density around it will change according to the density profile of the minihalo, Eq.~(\ref{rh}). Considering an impact parameter ranging as $0.01 R_{\text{halo}} \lesssim b \lesssim R_{\text{halo}}$, the typical crossing time is $\mathcal{O}(10^{11}\,\text{s})$ for the minihalo mass range of interest. Figure~\ref{Plot5b} (right) shows the contour plot of the crossing time in the parameter space of $(M_{\text{halo}}/M_{\text{PBH}}, b/R_{\text{halo}})$ at the local neighborhood. The smaller the impact parameter or the larger the minihalo mass in units of the central PBH mass, the larger the crossing time as expected. 
 
\subsection{Estimates for radio telescope sensitivity } 
 We estimate the current and projected sensitivity from radio emission during a NS-dressed PBH encounter. The minimum detectable flux for a given radio telescope reads as~\cite{Safdi:2018oeu}
 \begin{align}
 S_{\text{min}} &=\text{SNR}_{\text{min}} \frac{\text{SEFD}}{\eta_s \sqrt{2 \Delta B \Delta t_{\text{obs}}}}\,,\\\label{radiometereq}
 & \sim 220\, \mu\text{Jy}\left( \frac{\text{SNR}_{\text{min}}}{5} \right)\left( \frac{\text{SEFD}}{10\,\text{Jy}} \right) \left(  \frac{0.9}{\eta_s}\right) \left( \frac{1\,\text{kHz}}{\Delta B} \right)^{1/2} \left( \frac{1\,\text{yr}}{\Delta t_{\text{obs}}} \right)^{1/2}
 \end{align}where $\text{SNR}_{\text{min}}$ is the minimum signal-to-noise ratio,  $\text{SEFD} \equiv 2 k_{\text{B}} T_{\text{sys}}/A_{\text{e}}$ is the system equivalent flux density (where $T_{\text{sys}}$ and $A_{\text{e}}$ are the system temperature and the effective area,  respectively), $\Delta B$ is the bandwidth, $\eta_s$ is the system efficiency, and $\Delta t_{\text{obs}}$ is the observation time. 
 
 The Expanded Very Large array (EVLA) located in New Mexico  comprises 27 independent antennas and covers frequencies from $1\, \text{GHz}$ to $50\, \text{GHz}$. The antenna SEFD is  $\mathcal{O}(100)\, \text{Jy}$ with the minimum in the X-band (central frequency: 10 GHz) and maximum in the Q-band (central frequency: 45 GHz). The antenna SEFD in each band is reported in Table 1 of Ref.~\cite{2011ApJ...739L...1P}. For all antennas measurement, we scale the minimum detectable flux by the total number of antennas $N$, e.g. $S_{\text{min}}/\sqrt{N(N-1)}$ in Eq.~(\ref{radiometereq}) (see Sec. 3 in~\cite{EVLA}).
 The Square Kilometre Array (SKA) phase 1 to be constructed during the period 2018-2023 can cover frequencies from 50 MHz to 350 MHz (SKA-low frequencies) and from 350 MHz to 13.8 GHz (SKA-mild frequencies). The (all antennas) SEFD %of all antennas 
 for the SKA1-mid array is $\mathcal{O}(1)\, \text{Jy}$ being maximised in the band 1 (central frequency: $0.7\,\text{GHz}$) %reaching a value of  $ 2\times1380/779\, \text{Jy}$ 
 and minimised in the bands 2-3 (central frequency: $2\,\text{ GHz}$). %with a value $2\times 1380/1309\, \text{Jy}$.  
 The (all antennas) SEFD in each band is reported in Table 7 of Ref.~\cite{SKAGuide}. 
 From now on, we assume that the optimized bandwidth in Eq.~(\ref{radiometereq}) matches the signal bandwidth so that
 $\Delta \nu = \Delta B $.

 Regarding the NSs properties, a Monte Carlo-based population synthesis is used by Faucher and Kaspi ~\cite{FaucherGiguere:2005ny} to model the birth properties and time evolution of pulsars in the Parkes and Swinburne Multibeam surveys. While the magnetic field at the NS pole follows a log-normal distribution ranging as $\sim\left(10^{12}-10^{14}\right)\text{G}$, the spin period follows a normal distribution ranging as $\sim \left(0.04 - 10\right)\text{s}$ (see Figure 6 in Ref.~\cite{FaucherGiguere:2005ny}). As discussed in Refs.~\cite{Pshirkov:2007st, Hook:2018iia}, potentially good targets for axion detection via resonant conversion are the so-called Magnificent Seven, which corresponds to a group of $\sim 7$ nearby isolated NSs holding strong magnetic fields ($\sim 10^{13}\,\text{G}$), long spin periods ($\sim 5\,\text{s}$), short distances from the Earth (less than $500\,\text{pc}$), and absence of non-thermal emission and observed radio emission~\cite{Voges:1999ju, Kaplan:2008qn, Kaplan:2009ce}. From Eq.~(\ref{DPDOmeganoangle}), we see that such NS properties would enhance the flux during NS-dressed PBH encounters. In addition,  while the lack of non-thermal emission allows us to assume the validity of the Goldreich-Julian model within the NS magnetosphere, the lack of radio pulsar-like emissions allows us to estimate the projected sensitivity to the axion-photon coupling neglecting the NS background radiation~\cite{Hook:2018iia}. Within this group, we find the isolated NS RX J086.4-4123 which we mentioned before~\cite{Kaplan:2009ce}. 
 
\begin{figure}[H]
\centering
\includegraphics[width=11 cm]{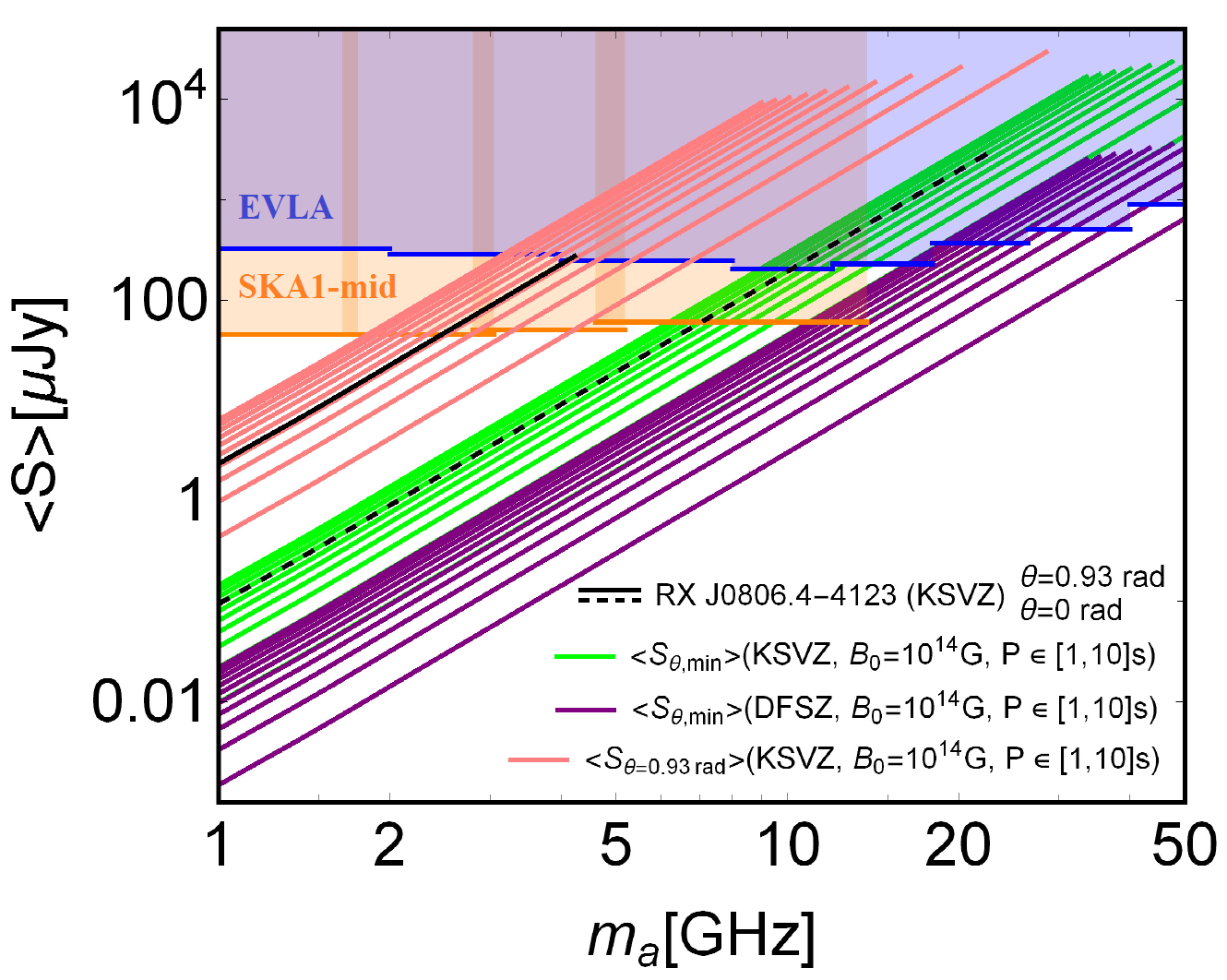}
\includegraphics[width=11.55 cm]{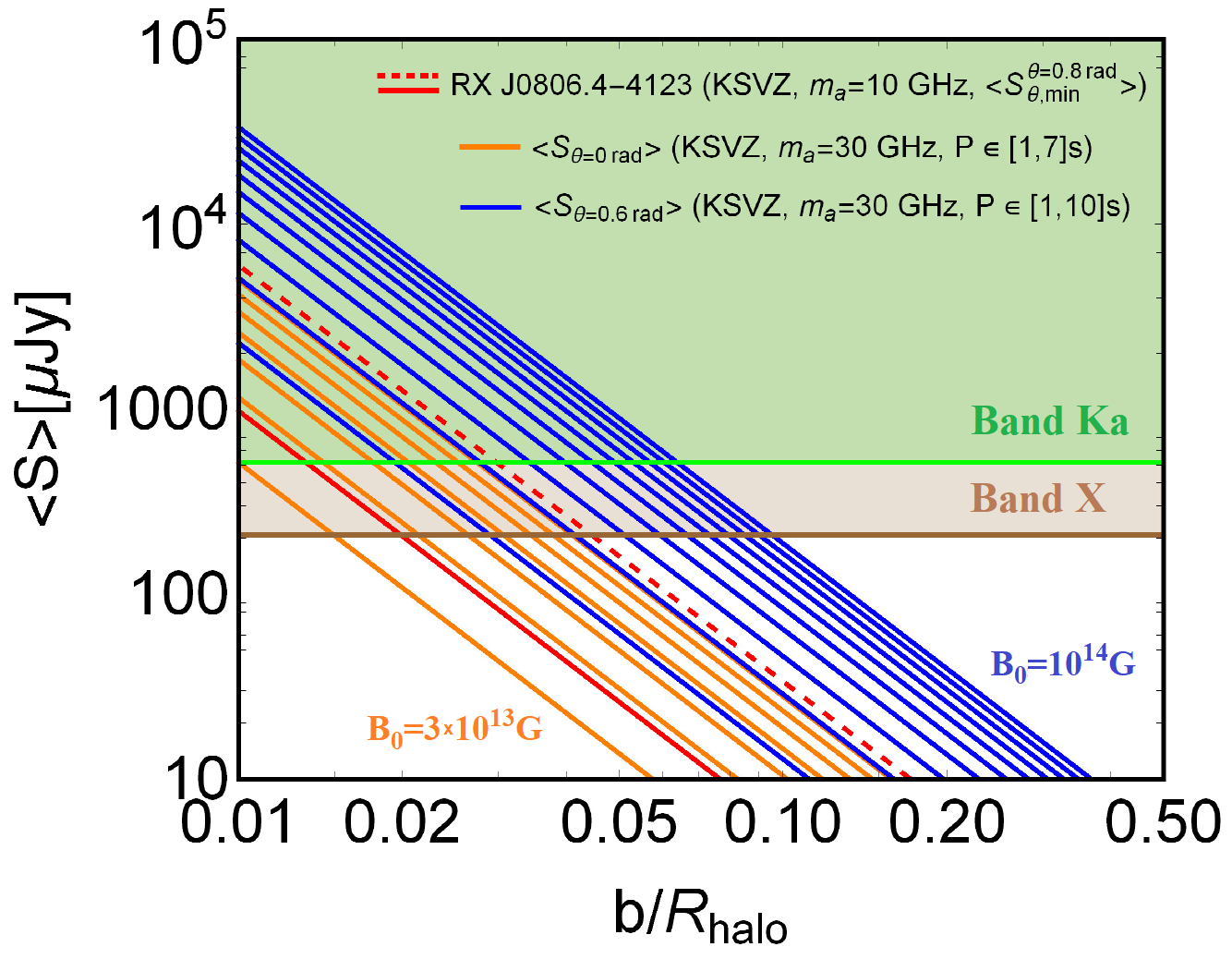}
\caption{ Here $d = 250 \,\text{pc}$, $\Delta B = \Delta_{\nu}=1\,\text{kHz}$, $M_{\text{NS}}= M_{\odot}$, $M_{\text{halo}} = 100\, M_{\text{PBH}}$, $\Delta t_{\text{obs}} = 1\,\text{ yr}$, $\text{SNR}_{\text{min}}=5$, and $\eta_s = 0.92$. (Top)  $\langle S \rangle$ for the KSVZ and DFSZ  models for $B_0 = 10^{14}\,\text{G}$ and  $\theta =  \{\pi/2\,, 0.93\,\}\text{rad}$. We assume the NS is crossing the zone within the minihalo at which the flux is maximum for $b = 0.01R_{\text{halo}}$. Each group of 10 curves (pink, green, purple) is obtained  for P from 1 s (lowermost curve) to 10 s (uppermost curve).  The shaded blue (orange) band is the detectability zone for EVLA (SKA1-mid). (Bottom) $\langle S \rangle$ for the KSVZ model and $m_a = 30\, \text{GHz}$ as function of $b/R_{\text{halo}}$. The group of 10 blue (7 orange) curves refers to $B_0 = 10^{14}\,\text{G}$, $\theta=0.6\, \text{rad}$ ($B_0 = 3\times 10^{13}\,\text{G}$,  $\theta=0\, \text{rad}$) and are obtained for P from 1 s to 10 s (to 7 s). The lowermost  (uppermost) curve refer to the minimum (maximum) spin period in each group. The green  (brown) shaded region is the detectability zone for EVLA Band Ka (X). We have added in both panels results for RX J0806.4-4123 using different $\theta$ angles and axion masses.}
\label{Plot6}
\end{figure}

 Figure~\ref{Plot6} (top panel) shows the estimated mean spectral flux density of one typical NS-dressed PBH encounter at $d=250\, \text{pc}$. We have taken
 $\Delta t_{\text{obs}}= 1 \text{year}$,  $B_{0} = 10^{14}\,\text{G}$, $M_{\text{NS}} = M_{\odot}$, $M_{\text{halo}} = 100\, M_{\text{PBH}}$, $\Delta B = 1\,\text{kHz}$, $\text{SNR}_{\text{min}} = 5$, and a system efficiency $\eta_s=0.92$~\footnote{The performance report of the EVLA states that its efficiency is at least 0.92~\cite{EVLA}.}. We are assuming that the NS is crossing the minihalo zone at which the flux is maximum for an impact parameter $b=0.01 R_{\text{halo}}$, i.e. in the middle of the observing period $\Delta t/2$ the distance $\tilde{R}_{\rm halo}$ is minimised, see Fig. 7 (left).  
 The zones for detection for the SKA1-mid array~\cite{SKAGuide} and EVLA~\cite{2011ApJ...739L...1P}  are shown in the shaded orange and blue bands, respectively. Each group of 10 flux curves is obtained by varying the NS spin period from 1 s (lowermost curve) to 10 s (uppermost curve). We have shown
 the particular cases for the minimum mean flux associated with the KSVZ (green curves) and DFSZ (purple curves) axion models and the mean flux at $\theta = 0.93\,\text{rad}$ for the KSVZ model (pink curves). \textcolor{black}{In addition, the solid (dashed) black line shows the estimated mean spectral flux density of the particular NS RX J0806.4-4123, assuming the KSVZ model and a polar angle $\theta = 0.93 \,\text{rad}$ (0\,\text{rad}).
 Generally speaking,} for sufficiently large axion mass, we see that the signal associated with both axion models should be detectable by both SKA-1 mid and EVLA telescopes \textcolor{black}{when we consider $\Delta B = 1\, \text{kHz}$}. 
 In the bottom panel of Fig.~\ref{Plot6}, the effect of the impact parameter is analyzed by taking otherwise the same parameters as above but using just the KSVZ model and fixing the axion mass to 30 GHz. The group of 10 blue (7 orange) curves are obtained by taking 
 $B_0 = 10^{14}\,\text{G}$ and $\theta= 0.6\,\text{rad}$ ( $B_0 = 3\times10^{13}\,\text{G}$ and $\theta=0\,\text{rad}$) and varying the NS spin period from 1 s to 10 s (1 s to 7 s). \textcolor{black}{In addition, the solid and dashed red lines show the particular case of RX J0806.4-4123, assuming the KSVZ model, $m_a = 10\, \text{GHz}$, and  $\theta= \pi/2$ and $0.8$ rad, respectively. The green (brown) shaded region corresponds to the detectability zone for EVLA Band Ka (X), which is sensitive to the spectral line associated with $m_a = 30\,\text{ GHz}$ ($10\,\text{GHz}$).  Generally speaking,} the smaller the impact parameter, the larger the flux as the NS crosses inner regions of the minihalo with higher densities. The axion detection is favored for $b \sim \mathcal{O}(0.01) R_{\text{halo}}$  for the predefined observation time and bandwidth. 
 
\begin{figure}[H]
\centering
\includegraphics[width=11.55 cm]{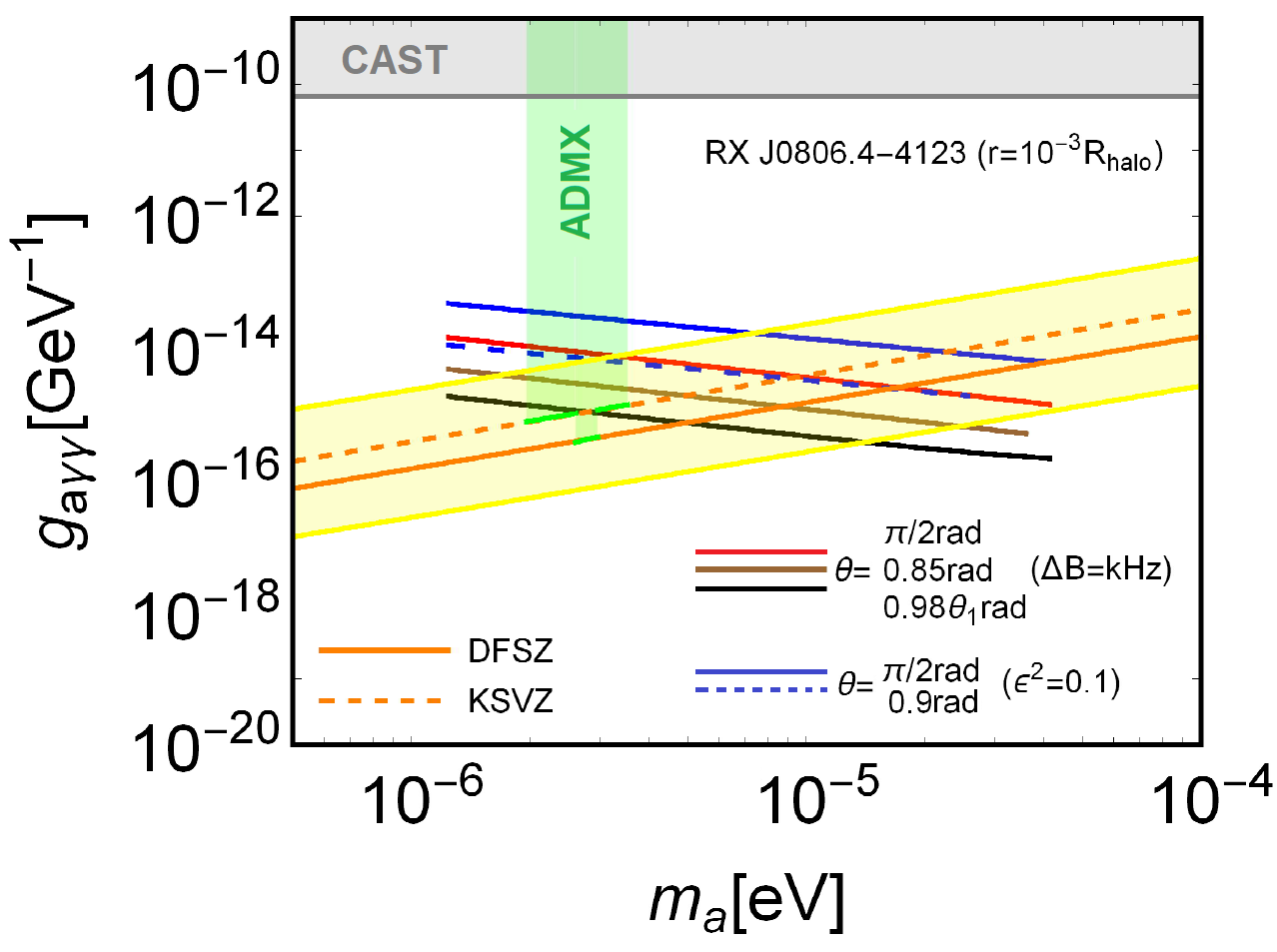}
\caption{ Projected sensitivity to $g_{a\gamma\gamma}$ as a function of the axion mass $m_a$ for $\Delta t_{\text{obs}}=100\, \text{hr}$, $\eta_s=0.5$, $\text{SNR}_{\text{min}}=1$, and $\text{SEFD} = 2\, \text{Jy}$ for an Arecibo-like telescope. The shaded yellow band is the QCD axion parameter space, while the dashed (solid) orange line is the particular KSVZ (DFSZ) model. Exclusion regions from CAST~\cite{Anastassopoulos:2017ftl} and ADMX~\cite{PhysRevLett.120.151301} are shown by the gray and green shaded bands, respectively. The red, brown, and black solid lines correspond to the projected sensitivity for the isolated NS RX J0806.4-4123 using $\theta = (\pi/2, 0.85, 0.98\theta_1)$\,rad, respectively, and  $\Delta B= \Delta \nu = 1\, \text{kHz}$. The  solid and dashed blue lines 
are calculated using $\Delta \nu$ from Eq.~(\ref{B}) with $\epsilon^2=0.1$ and $\theta = \pi/2$ and $0.9$ rad, respectively.
For all cases, we are assuming a NS passing through a dressed PBH ($M_{\text{halo}}=10^2 M_{\text{PBH}}$) at a distance $r=10^{-3}R_{\text{halo}}$ from its central PBH.}
\label{Plot7}
\end{figure}

 Figure~\ref{Plot7} shows the projected sensitivity to the axion-photon coupling constant as a function of the axion mass for the isolated NS RX J0806.4-4123 during one encounter with a dressed PBH with $M_{\text{halo}} = 10^2 M_{\text{PBH}}$. We use a single-dish Arecibo-like telescope with a typical $\text{SEFD} = 2\,\text{Jy}$ in the frequency range from $300\, \text{MHz to}\, 10\, \text{GHz}$, $\eta_s = 0.5$, and $\text{SNR}_{\text{min}}=1$. We are estimating the corresponding mean spectral flux density as $\langle S(\theta) \rangle \approx S(\theta)$ for $\Delta t = \Delta t_{\text{obs}} = 100\,\text{hr}$ and $\rho_a(r_c) \approx \rho_a(r)$, where $r=10^{-3}R_{\text{halo}}$ is the radial distance of the NS from the central PBH.
 The gray and green shaded bands correspond to the exclusion regions from the CAST solar axion experiment~\cite{Anastassopoulos:2017ftl}
 and the ADMX haloscope collaboration~\cite{PhysRevLett.120.151301}, respectively. The yellow shaded region is the QCD axion parameter space, where we have indicated with a dashed and solid orange lines the particular KSVZ and DFSZ models, respectively.
 For $\Delta B  = 1\,\text{kHz}$, the red, brown, and black  solid lines show the projected sensitivity for RX J0806.4-4123 assuming a polar angle $\theta = (\pi/2, 0.85, 0.98\theta_1)\,\text{rad}$, respectively, where $\theta_1=\theta_1(m_a,P,B_0)$ is calculated using Eq.~(\ref{theta3}).  For $\Delta B = \Delta \nu$ calculated using Eq.~(\ref{B}), the solid and dashed blue lines show  the projected sensitivity for RX J0806.4-4123 assuming $\epsilon^2 = 0.1$ and a polar angle $\theta = \pi/2$ and $0.9\,\text{rad}$, respectively.  
 We see that most part of the sensitivity curves are within the QCD axion parameter for different angles of observation and $\epsilon^2 \leq 0.1$.


\section{Discussion and Conclusion}

In this paper, we have discussed for the first time a novel way to detect the QCD axion by means of transient radio signatures from resonant axion-photon conversion during encounters of NSs and axion minihalos around PBHs in the Milky Way. In the scenario  where the  PQ symmetry is broken before (or even during) the inflation and a small fraction of DM is  composed of PBHs, they will unavoidably acquire minihalos from the axion background.
 Thus, dressed PBHs will end up in the Milky Way halo at $z\sim 6$ to undergo later different levels of disruption.
 Mainly motivated by the LIGO-Virgo gravitational waves detection~\cite{Abbott:2017vtc, Sasaki:2016jop} and the recent NANOGrav results~\cite{Arzoumanian:2020vkk}, as a Benchmark model, we take $f_{\text{PBH}}=10^{-3}$ as the initial fraction of (naked) PBHs with masses $M_{\text{PBH}} = \mathcal{O}(1-10)\,M_{\odot}$. For such range in PBH masses, the main sources of disruption are tidal forces due to
the mean field potential of the Milky Way and the gravitational  potential during disk crossing.
Under the simplification of circular orbits around the Galactic center, we estimate the mass loss of minihalos depending on their Galactocentric radii. Assuming an initial Gaussian minihalo mass distribution, we 
conclude that the mass loss is negligible for radii $R \gtrsim 13 \,\text{kpc}$, but as we approach the Galactic center the loss increases reaching up to $\sim 90\%$ of the original mass for $M_{\text{halo}} \sim (135-270) M_\text{PBH}$ at $R \sim (0.1-2)\,\text{kpc}$.
The heavier the minihalo in units of the central PBH and the smaller its orbital radius, the larger the mass loss due to disruption in the Milky Way.

 Taking into account the disruption effects that act on the dressed PBHs,
we estimate the differential encounter rate between NSs and dressed PBHs as function of the 
Galactocentric
 radius. In the local neighborhood, we find $d\Gamma_{\text{NS-PBH}}/dr \sim 2\times 10^{-7}\,\text{kpc}^{-1}\text{s}^{-1} (M_{\text{PBH}}/M_{\odot})^{-1/3}$. As we approach to the inner parts of the Milky Way the number of NSs and dressed PBHs increases but the  mass  distribution  of  dressed  PBHs  also moves  towards  lighter minihalo  masses (and correspondingly smaller minihalo diameters). 
The two effects
tend to compensate each other at distances below
leading to a differential encounter rate of $\sim 10^{-7}\,\text{kpc}^{-1}\text{s}^{-1}$. 
At such small distances our assumption of re-viralization of the minihalos between disk crossings starts to break down which is expected to lead to certain overestimation of the differential encounter rate. This should not affect our final results, however, as the small distance regime gives only a subleading contribution to the total encounter rate.
The encounter rate scales with the PBH mass as $\Gamma_{\text{NS-PBH}} \sim (M_{\text{PBH}}/M_{\odot})^{-1/3}$, so that the larger the central PBH mass the smaller is the encounter rate between NSs and dressed PBHs. 
The minihalo radius grows 
as $\sim (M_{\text{PBH}}/M_{\odot})^{1/3}$ but the number density of  dressed PBHs in the Milky Way halo decreases as $\sim (M_{\text{PBH}}/M_{\odot})^{-1}$. Therefore, when $M_{\text{PBH}}$ is increased, the decrease of the number of dressed PBHs available for encounters dominates over the increase of the collision cross section.  
As the total encounter rate, integrated up to 
$R\geq 100 \,\text{pc}$, we find 
$\mathcal{O}(10^{-2})\,\text{day}^{-1}$
 in the
PBH mass range of our interest.
Our estimates for the rate of NS-dressed PBH encounters  could be improved
%needs to be refined 
by  including
eccentric orbits around the Galactic Center in the study and/or performing numerical simulations to statistically track the evolution of the minihalo initial mass distribution in the Milky Way halo, we leave these tasks for a future work. 

By estimating the mean spectral flux density produced by the NS-dressed PBH interaction and comparing it with the sensitivity of current and prospective radio telescopes, we show that the transient radio signals should be detectable on the Earth under suitable conditions. Because of the power radiated per unit solid angle scales as $dP/d\Omega(\theta = \pi/2) \sim (\rho_{\text{halo}}) (m_a)^{4/3} (B_0)^{5/6} (P)^{7/6}  (g_{a\gamma\gamma})^{2}$, the signal is the stronger the smaller the impact parameter, and the larger the magnetic field, spin period, axion-photon coupling constant and axion mass. 

\textcolor{black}{ Interesting targets for possible detection
%Suitable targets for detection 
would be nearby isolated NSs having long spin periods,
strong magnetic fields, and absence of non-thermal emission and pulsar activity. We use as a  particular example, the isolated NS RX J0806.4-4123, which is part of the Magnificent Seven group~\cite{Voges:1999ju, Kaplan:2008qn, Kaplan:2009ce}. Based on the sensitivity of current and prospective radio telescopes, we show in Figs.~\ref{Plot6} and \ref{Plot7} that RX J0806.4-4123 could
%should 
be detectable 
on the Earth if it undergoes a close encounter with a typical dressed PBH. Since the mean spectral flux density is proportional to the axion density at the conversion radius and the minihalo density quickly increases as we approach the inner shells, the minimum observation time required for detection significantly decreases if we assume smaller impact parameters. In Fig.~(\ref{Plot6}) we show that the transient radio signal coming from a RX J0806.4-4123-dressed PBH encounter could be detectable on the Earth assuming EVLA and SKA1-mid sensitivities, $b \sim \mathcal{O}(0.01) R_{\text{halo}}$, a kHz-bandwidth and 1 year of observation. In Figure~\ref{Plot7} we show that the projected sensitivity to the axion-photon coupling constant, assuming an Arecibo-like telescope and 100 hours of observation time, is mostly within the QCD axion parameter space, including the particular KSVZ and DFSZ models, for a bandwidth of the order of $(1-10^2)\,\text{kHz}$. Note that our sensitivity results for  RX J0806.4-4123 with a bandwidth
$\sim 10^2\,\text{kHz}$ are more than 3 orders of magnitude stronger than those of Ref.~\cite{Battye:2019aco}, which is currently one of the state-of-the-art estimates. This is because the axion minihalo density at a distance $r=10^{-3}R_{\text{halo}}$ from the central PBH is about 8 orders of magnitude higher than the local DM density and the sensitivity of the axion-photon coupling constant scales as $g_{a\gamma\gamma} \sim 1/\sqrt{\rho_a(r_c)}$. %Indeed, re-scaling for the bandwidth and the distance from the source, our results are comparable with their results assuming resonant axion-photon conversion of a typical NS close to Sagittarius $A^*$.%
}

Motivated by the enhancement in the signal via the enhancement in the axion DM background, there has been much interest in the literature for studying the signal coming from  axion stars located close to the Galactic center, such as the magnetar SGR J1745-2900. This astrophysical object is located at about 0.1 pc from the Galactic center and holds a strong magnetic field at poles $B_0 \approx 1.6 \times 10^{14}\,\text{G}$, and a long spind period $P\sim 3.76\, \text{s}$~\cite{Kennea:2013dfa, Mori:2013yda, Shannon:2013hla, Eatough:2013nva}. However, such kind of an environment is not the ideal in our setup since the strong tidal forces coming from the Milky Way's central supermassive black hole would remove most of the DM shells from minihalos.

Extragalactic encounters of NSs with dressed PBHs may potentially  also be detectable on the Earth if we consider close encounters which lead to a final merger.   
In such a scenario, we could 
have a multi-messenger probe of the QCD axion via the observation of both  the axion photon conversion signal and  gravitational waves from PBH-NS binary inspirals, as studied in detail in Ref.~\cite{Edwards:2019tzf} for the case of intermediate mass black holes with axion DM spikes.

Even though locations of most of the NSs are not know,  \textcolor{black}{we consider that a good starting point for searches should be along the Milky Way disk and towards the Galactic center (but sufficiently far away to avoid sizeable minihalo disruption) based on the NSs sub-population on the disk and bulge.} On the other hand, resolving the signal from radio foregrounds is expected to be a general challenge for such searches involving large fields of view but analysis of this topic goes beyond the scope of the current work.

Recently, a study of radio signals coming from the axion-photon conversion during NS-axion minicluster encounters was performed in Ref.~\cite{Edwards:2020afl}. According to the sensitivity of current and prospective radio telescopes, they showed that the QCD-axion should be detectable 
assuming an axion mass of $\sim 5\,\text{GHz}$ and a  power-law density profile for axion miniclusters with large mean density, $\sim(10^4-10^8) M_{\odot}/\text{pc}^3$. If axion miniclusters are assumed to have a NFW density profile, the associated emitted flux  would not be large enough to be detectable (for a discussion about the axion minicluster density profile in the context of N-body simulations see, for example, Ref.~\cite{Eggemeier:2019khm}). If the PQ symmetry breaking occurs before or during the inflation the axion fluctuation is too small to form axion miniclusters, but if the PQ symmetry is broken after the inflation, large density fluctuation of  the axion will be produced. However, it seems uncertain how large axion fluctuations are produced after the PQ symmetry breaking. Our scenario
on the other hand assumes the presence of PBHs which is certainly uncertain, although 
the LIGO-Virgo observations and recent NANO-Grav results may provide some suggestive hints of 
their existence. 

To conclude, our results give rise to the striking prediction of a transient characteristic line-like emission which would need to be confirmed via dedicated search strategies.
A more careful analysis about the final bandwidth of the radio signal and model for the conversion process happening in the NS magnetosphere should be done. We leave these tasks for future work.

\section*{Acknowledgments}
 This work was supported by the Academy of Finland grant 318319.
 E. D. S. thanks Martin Stref and Paul McMillan for discussions about the Milky Way Galactic model.  T. T. Y. is supported in part by the China Grant for Talent Scientiﬁc Start-Up Project and the JSPS Grant-in-Aid for Scientiﬁc Research Grants No. 16H02176, No. 17H02878, and No. 19H05810 and by World Premier
International Research Center Initiative (WPI Initiative), MEXT, Japan.

\bibliographystyle{JHEP}
\bibliography{QCDAxionNS}
\end{document}